\documentclass[twocolumn]{aastex631} 

\def\apj{{ ApJ}}
\def\apjl{{ ApJL}}
\def\apjs{{ ApJS}}
\def\apss{{ Ap\&SS}}
\def\aap{{ A\&A}}
\def\aj{{ AJ}}
\def\mnras{{ MNRAS}}

\def\nat {{ Nature}}

\def\ssr{{ Space Sci. Rev.}}

\def\PRE{{ Phys. Rev. E}}
\def\prd{{ Phys. Rev. D}}
\def\PRL{{ Phys. Rev. Lett}}

\def\jcap{{ JCAP}}

\def\pasp{{ PASP}}

\newcommand{\be}{\begin{equation}}
\newcommand{\ee}{\end{equation}}
\newcommand{\bea}{\begin{eqnarray}}
\newcommand{\eea}{\end{eqnarray}}

\shorttitle{Evolution of Highly Magnetic White Dwarfs}
\shortauthors{Bhattacharya et al.}

\begin{document}

\title{Evolution of highly magnetic white dwarfs by field decay and cooling: theory and simulations
}
 
\author{Mukul Bhattacharya}
\affiliation{Department of Physics, The Pennsylvania State University, University Park, PA 16802, USA; mmb5946@psu.edu}

\author{Alexander J. Hackett}
\affiliation{Institute of Astronomy, University of Cambridge, Madingley Road, Cambridge CB3 0HA; ajh291@cam.ac.uk}

\author{Abhay Gupta}
\affiliation{School of Physics, The University of New South Wales, Sydney, New South Wales 2052, Australia; abhayrgupta06@gmail.com}

\author{Christopher A. Tout}
\affiliation{Institute of Astronomy, University of Cambridge, Madingley Road, Cambridge CB3 0HA; cat@ast.cam.ac.uk}

\author{Banibrata Mukhopadhyay}
\affiliation{Department of Physics, Indian Institute of Science, Bangalore 560012, India; bm@iisc.ac.in}

\begin{abstract}
We investigate the luminosity suppression and its effect on the mass--radius relation as well as cooling evolution of highly magnetised white dwarfs. 
Based on the effect of magnetic field relative to gravitational energy, we suitably modify our treatment of the radiative opacity, magnetostatic equilibrium and degenerate core equation of state to obtain the structural properties of these stars. 
Although the Chandrasekhar mass limit is retained in the absence of magnetic field and irrespective of the luminosity, strong central fields of about $10^{14}\, {\rm G}$ can yield super-Chandrasekhar white dwarfs with masses $\sim2.0\, M_{\odot}$. 
Smaller white dwarfs tend to remain super-Chandrasekhar for sufficiently strong central fields even when their luminosity is significantly suppressed to $10^{-16}\ L_{\odot}$. Nevertheless, owing to the cooling evolution and simultaneous field decay over $10\ {\rm Gyr}$, the limiting masses of small magnetised white dwarfs can fall to $1.5\ M_{\odot}$ over time. However the majority of these systems still remain practically hidden throughout their cooling evolution because of their high fields and correspondingly low luminosities. 
Utilising the stellar evolution code \textsc{stars}, we obtain close agreement with the analytical mass limit estimates and this suggests that our analytical formalism is physically motivated. 
Our results argue that super-Chandrasekhar white dwarfs born due to strong field effects may not remain so forever. This explains their apparent scarcity in addition to making them hard to detect because of their suppressed luminosities.
\end{abstract}


\section{Introduction}
\label{sec1}

Observations of more than a dozen overluminous Type Ia supernovae (SNe Ia; \citealt{Howell2006}; \citealt{Scalzo2010}) strongly indicate the existence of their very massive progenitors  
with masses $M > 2\, M_{\odot}$. SNe Ia are widely studied because of their utility as standard candles to estimate cosmological distances. 
While binary evolution of accreting/rapidly differentially rotating white dwarfs (WDs; \citealt{Hachisu86,YL2004}) have already been used to explain such high mass progenitors, none of these models could account for progenitor masses as high as $2.8M_\odot$ that are inferred from the observations. 
Therefore, an alternate but equally exciting proposal relates to the highly magnetized super-Chandrasekhar white dwarfs (here after B-WDs). In addition to being the super-Chandrasekhar mass progenitors of overluminous SNe Ia, B-WDs have also been proposed as promising candidates for soft gamma-ray repeaters (SGRs) and anomalous X-ray pulsars (AXPs) that have significantly lower magnetic fields and ultraviolet luminosities \citep{MR2016}. 

Previous studies \citep{DM2012,DM2013,SM2015} already showed that strong magnetic fields can appropriately modify the equation of state (EoS) of the electron degenerate matter to yield super-Chandrasekhar WDs with $M \approx 2.6\, M_{\odot}$, 
with or without sufficiently rapid rotation. Interestingly, a misalignment between the rotation and magnetic axes can generate, along with dipole radiation, 
significant 
gravitational radiation which can be detected by space-based gravitational wave detectors leading to a direct detection of super-Chandrasekhar WDs \citep{KM2019}. 
Observations confirm that magnetised WDs  are indeed more massive than other non-magnetised WDs \citep{FMG2015}. Furthermore, data from the Sloan Digital Sky Survey (SDSS) suggest that, in addition to having marginally higher masses, magnetised WDs also span a similar effective temperature range as their non-magnetic counterparts \citep{Van2005}. 

The effect of strong magnetic fields on the WD mass--radius relation has already been explored in some detail previously by our group, for both Newtonian \citep{DM2012} and general relativistic formalisms \citep{DM2014,SM2015,DM2015}. These investigations were carried out for various different magnetic field configurations and were in good agreement with the results from independent studies that indicate the existence of super-Chandrasekhar WDs \citep{Boshkayev2013,FS2015,Carvalho2018}. \citet{Otoniel2019} recently studied potential matter instabilities for these B-WDs in a general relativistic framework, particularly in relation to the pycnonuclear and electron capture reactions. While they found that the limiting mass for non-rotating B-WDs is about $2.14 M_{\odot}$ and potentially larger if rotation is considered, the pycnonuclear reactions are likely to destabilise the star once the central density exceeds about $10^{10}\ {\rm g\ cm^{-3}}$. 

It should be noted that magnetised WDs have many important implications apart from their apparent link to overluminous SNe Ia and, hence, their other properties should also be explored \citep{MR2016,MRB2017,MDRSB2017}. Apart from increasing the limiting mass of WDs, strong magnetic fields can also influence the thermal properties, such as luminosity, temperature gradient and cooling rate of the star. \citet{MBh2018} explored the luminosities and cooling rates of B-WDs with the theoretical model proposed by \citet{ST1983} for non-magnetic WDs. Assuming that the interface properties are similar for both B-WDs and non-magnetised WDs for a given stellar age and a non-zero temperature gradient across the surface layers, they showed that the luminosity for B-WDs can be suppressed significantly up to about $10^{-9}\ L_{\odot}$ for large fields $B \gtrsim 10^{12}\, {\rm G}$. They also computed the cooling rates for these B-WDs with suppressed luminosities and showed that their cooling evolution is significantly slowed down at large $B$. Indeed, \citet{Valyavin2014} analysed the optical data for cool WD 1953-011 and found that strong fields suppress convection over the entire B-WD surface and thereby attenuate the cooling rate. 

The initial exploration into the cooling of non-magnetised WDs started with attempts to model the degenerate core as the primary source of thermal energy, which is then radiated away as the observed luminosity from the surface layers as the star gradually evolves over time \citep{Mestel1952,MR1967}. \citet{TY1996}  calculated the cooling curves for low-mass WDs starting from a luminous star and evolving to the crystallisation stage after about $10\, {\rm Gyr}$, while \citet{FBB2001} discussed some
limitations of the \citet{Mestel1952} model in the context of WD cosmochronology. However it is important to note that these studies either did not consider the effects of strong fields or assumed that the underlying fields were just too weak to have any practical effects on the WD cooling evolution. 

Recently, \citet{Gupta2020} revisited the physics of luminosity suppression and the mass--radius relation in the context of B-WDs. In contrast to \citet{MBh2018}, they relaxed the assumption of fixing interface parameters and a preassigned mass or radius across all B-WDs and non-magnetised WDs. They further extended their analysis to compute the radial profiles of the B-WD thermal properties for both the non-degenerate surface layers and the electron-degenerate B-WD cores. However they ignored the effect of strong fields on the EoS of the degenerate core electrons as well as the correction to the total B-WD mass by general relativistic effects. 

Here, in a considerably more generalised framework, we model the B-WD structure properties from the centre to the surface by solving the magnetostatic equilibrium, mass conservation and photon diffusion equations simultaneously. We investigate the effect of the temperature gradient (directly related to the luminosity) on the mass--radius relation by considering both radiative and convective cooling. While the total pressure is the sum of the contributions from the electron degenerate, ideal gas and magnetic pressures, the interface location is taken to be the radius where the degenerate pressure is roughly comparable to the ideal gas pressure. 

In order to distinguish between the weakly and strongly magnetised cases, we compare the relative energy densities in the magnetic and gravitational fields. Accordingly, we modify the radiative opacity, magnetostatic balance equation and the EoS for the electron-degenerate core. 
This paper is organised as follows. 
In Section 2, we provide a brief overview of the method used to obtain the radial profiles of the WD properties as well as the mass--radius relation for a given luminosity. 
In Section 3, we discuss the physical effects of the magnetic field for both the weak and strong field limits, based on a comparison with the gravitational energy of the B-WD. 
Subsequently, in Section 4, we evaluate the suppressed luminosity of strongly magnetised B-WDs, after including the effects of cooling evolution and magnetic field decay by dissipative processes over long timescales. 
Next, we explore a set of numerical models produced using the stellar evolution code \textsc{stars} in order to validate our analytical approach in Section 5 and finally conclude with a summary of the main results in Section 6.

\section{White dwarf structure properties}
In this section, we describe the physical considerations used to self-consistently obtain the structure properties of magnetised WDs. We formulate a method to solve the magnetostatic equilibrium, mass conservation and photon diffusion equations for a given luminosity and magnetic field configuration. We model the total pressure by including the contributions from the degenerate electron gas (dominant within the isothermal core), ideal gas (dominant in the surface layer) and magnetic pressures. The interface is defined to be at the WD radius where the contributions from the inner electron degenerate core and outer ideal gas pressures are equal. 
For our study, we consider radially varying magnetic fields that are realistic (see \citealt{Deb2021}). 
The presence of strong fields inside compact stars gives rise to additional pressure $P_B = B^2/8\pi$ as well as density $\rho_B = B^2/8\pi c^2$, where $B = \sqrt{\textbf{B}.\textbf{B}}$ is the strength of the magnetic field \citep{Sinha2013,MBh2018,MB2018b}.

\begin{figure*} 
\includegraphics[width=0.49\textwidth]{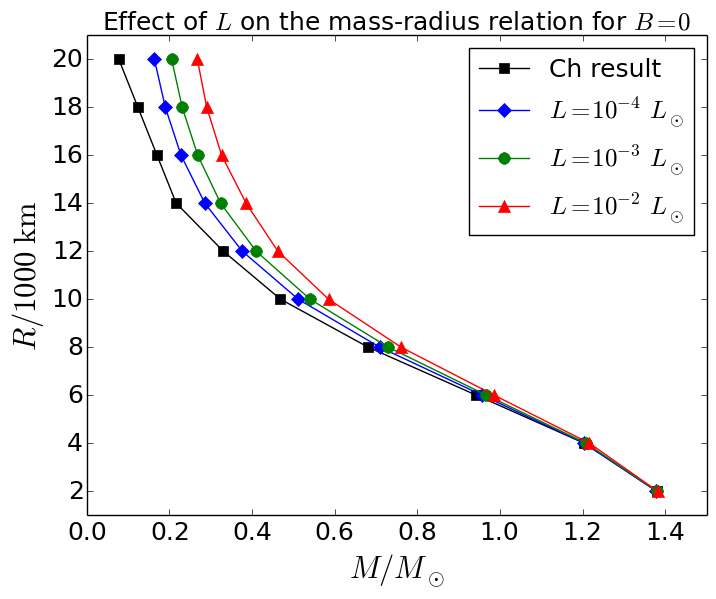} 
\includegraphics[width=0.49\textwidth]{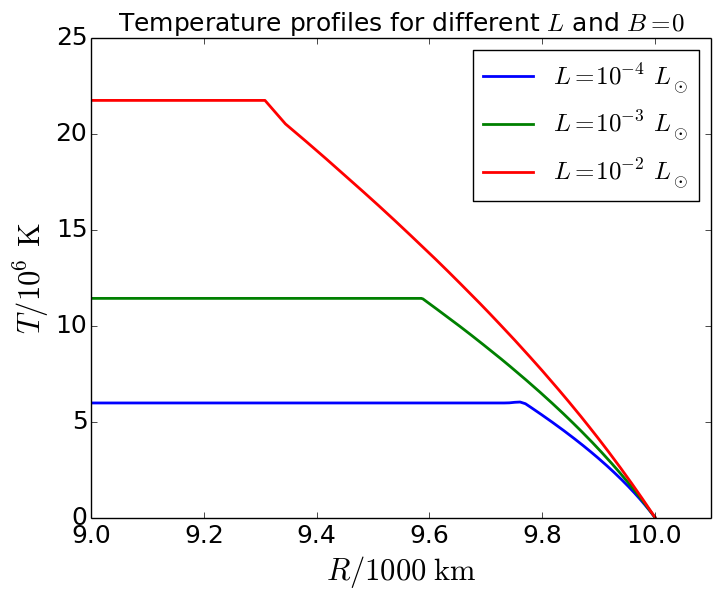} 
\caption{{\bf Left panel:} The effect of surface luminosity on the mass--radius relation of non-magnetised WDs is shown for the Chandrasekhar result (black squares), $L=10^{-4}\ L_{\odot}$ (blue diamonds),
$L=10^{-3}\ L_{\odot}$ (green circles) and $L=10^{-2}\ L_{\odot}$ (red triangles). 
{\bf Right panel:} The variation of temperature as a function of radius is shown for non-magnetised WDs with $L=10^{-4}\ L_{\odot}$ (blue curve),
$L=10^{-3}\ L_{\odot}$ (green curve) and $L=10^{-2}\ L_{\odot}$ (red curve).
} 
\label{fig1}
\end{figure*}

\begin{table*}
\caption{The effect of luminosity $L$ on the mass--radius relation for non-magnetised WDs. Here $M$ is the total mass and $R$ is the WD radius. The core temperature and density are $T_{\rm c}$ and $\rho_{\rm c}$ respectively, while $\rho_*$ and $R_*$ denote the interface density and radius.
}
\label{Table1}
\bgroup
\def\arraystretch{1.0}
\begin{tabular}{| c | c | c | c | c | c | c |}
\hline
\hline
\centering
$R/1000\ {\rm km}$ & $L/L_{\odot}$ & $T_{\rm c}/10^6 \ {\rm K}$ & $\rho_{\rm c}/10^6 \ {\rm g\ cm^{-3}}$ & $\rho_*/10^6 \ {\rm g\ cm^{-3}}$ & $R_*/1000\ {\rm km}$ & $M/M_{\odot}$ \\ \hline \hline
2.0 & $10^{-4}$ & 4.65 & 2215 & 0.0005 & 1.997 & 1.378 \\ 
    & $10^{-3}$ & 8.72 & 2243 & 0.0016 & 1.994 & 1.379 \\
    & $10^{-2}$ & 17.20 & 2286 & 0.0040 & 1.989 & 1.381 \\ \hline
6.0 & $10^{-4}$ & 5.00 & 23.41 & 0.0006 & 5.962 & 0.956 \\ 
    & $10^{-3}$ & 9.64 & 24.47 & 0.0016 & 5.924 & 0.967 \\
    & $10^{-2}$ & 18.49 & 26.07 & 0.0041 & 5.860 & 0.986 \\ \hline
12.0 & $10^{-4}$ & 6.82 & 0.827 & 0.0012 & 11.427 & 0.376 \\ 
    & $10^{-3}$ & 12.31 & 0.966 & 0.0021 & 11.171 & 0.410 \\
    & $10^{-2}$ & 23.04 & 1.819 & 0.0082 & 10.469 & 0.462 \\ \hline 
20.0 & $10^{-4}$ & 8.24 & 0.097 & 0.0015 & 16.208 & 0.164 \\ 
    & $10^{-3}$ & 15.22 & 0.136 & 0.0043 & 14.976 & 0.205 \\
    & $10^{-2}$ & 27.01 & 0.205 & 0.0093 & 13.542 & 0.268 \\ \hline \hline
\end{tabular}
\egroup
\end{table*}

We assume the B-WDs to be approximately spherical. The assumption of spherical symmetry is justified if the central field does not exceed about $10^{14}\, {\rm G}$ \citep{SM2015}, which is toroidally 
dominated, so the model equations characterising magnetostatic equilibrium, photon diffusion and mass conservation can be written within a Newtonian framework as
\begin{eqnarray} 
&\frac{{\rm d}}{{\rm d}r}(P_{\rm deg}+P_{\rm ig}+P_{B}) = -\frac{Gm(r)}{r^2}(\rho + \rho_B), \\
&\frac{{\rm d}T}{{\rm d}r} = -{\rm max}\left[\frac{3}{4ac} \frac{\kappa \rho}{T^3} \frac{L_r}{4\pi r^2}, \left(1 - \frac{1}{\gamma}\right)\frac{T}{P} \frac{{\rm d}P}{{\rm d}r}\right], \vspace{0.1in} \\
&\frac{{\rm d}m}{{\rm d}r} = 4\pi r^2 (\rho + \rho_B),
\end{eqnarray} 
where the magnetic tension terms are ignored for a radially varying $B$. In these equations, $P_{\rm deg}$ and $P_{\rm ig} = \rho kT/\mu m_p$ are the electron degeneracy pressure established by \citet{Chandrasekhar1935} and the ideal gas pressure respectively, $\rho$ is the matter density, $k$ is Boltzmann's constant, $T$ is the temperature, $\mu \approx 2$ is the mean molecular mass per electron, $m_p$ is the proton mass, $G$ is the Newton's gravitational constant, $m(r)$ is the mass enclosed within radius $r$, $a$ is the radiation constant, $c$ is the speed of light, $\kappa$ is the radiative opacity, $L_r$ is the luminosity at radius $r$, and $\gamma$ is the adiabatic index of the gas. 

The second term on the right-hand-side of equation (2) is the convective cooling contribution which can be dominant over the radiative photon cooling at some combinations of $T$ and $P = P_{\rm deg}+P_{\rm ig}+P_{B}$. Although strong central magnetic fields can potentially impede convection \citep{CM1992,Solanki2003}, a large fraction of the total energy flux can still be efficiently transported across the B-WD. Hence, we consider its effect on the stellar temperature profile in addition to the radiative cooling. The opacity in the surface layers of a non-magnetised WD can be approximated by the Kramers' formula, $\kappa = \kappa_0 \rho T^{-3.5}$, where $\kappa_0 = 4.34\times10^{24}Z(1+X)\ {\rm cm^2\ g^{-1}}$, and $X$ and $Z$ are the mass fractions of hydrogen and heavy elements (other than hydrogen and helium) in the stellar interior respectively.
Assuming helium WDs for our purpose here, we set the helium mass fraction to $Y=0.9$ and $Z=0.1$ for simplicity. 
The radiative opacity in the surface layers is primarily due to the bound-free and free-free transitions of electrons \citep{ST1983}. The radiation conduction typically dominates over the electron conduction in these regions and hence the same goes with the corresponding opacities \citep{PY2001}. However, in the presence of strong magnetic fields, the radiative opacity depends strongly on $B$, as we shall discuss in the next section.

It should be noted that there will be residual currents generated due to the variation of magnetic field within the star. However, the entire degenerate B-WD core is essentially isothermal due to its very large thermal conductivity, which also leads to the frozen magnetic flux. 
As a consequence, the heating ability is minimal given the near superconducting nature of the degenerate core and magnetic dissipation turns out to be negligible.

While a large number of B-WDs with surface fields up to $10^9\, {\rm G}$ or so have already been discovered by the Sloan Digital Sky Survey (SDSS; \citealt{Schmidt2003}), it is likely that the central fields are several orders of magnitude larger \citep{Fujisawa2012,DM2014,SM2015}. This is expected for residual field arising from the fossil field of the original star which had a stronger core field in addition to dynamo effects that can replenish those fields (see \citealt{PT2010,DMR2013,MST2021}. In order to capture the variation of field magnitude radially within the B-WD, here we adopt a profile used extensively to model magnetised neutron stars (NSs) and B-WDs \citep{Bandyopadhyay1997,DM2014,Deb2021}, 
\be
B\left(\frac{\rho}{\rho_0}\right) = B_{\rm s} + B_0 \left[1 - {\rm exp}\left(-\eta \left(\frac{\rho}{\rho_0}\right)^{\gamma}\right)\right].
\label{Bprof}
\ee
Here $B_{\rm s}$ is the surface magnetic field, $B_0$ is a fiducial magnetic field, $\eta$ and $\gamma$ are dimensionless parameters that determine how the magnetic field changes from the core to the surface. As $\rho \rightarrow 0$ close to the 
WD surface, $B \rightarrow B_{\rm s}$, whereas $\rho \rightarrow \rho_{\rm c}$ near the B-WD core that leads to $B \rightarrow B_0$. For our analysis here, we set $\rho_0 = 10^9\, {\rm g\, cm^{-3}}$, $\eta=0.8$ and $\gamma=0.9$ for all calculations, following \citet{MBh2018,MB2018b}. 
The profile in equation (\ref{Bprof}) essentially indicates the magnitude of field at  various density points within the star and hence radial coordinates.
Here we neglect effects such as offset dipoles and magnetic spots that can arise with more complex field configurations \citep{MM1999,Vennes2003}. 
Hereafter, we denote the magnetic field as $B=(B_{\rm s},B_0)$ in order to specify both the surface and core fields.

It should be noted that the model field profile given by equation (\ref{Bprof}) is not a unique choice and alternate profiles have been explored in the literature, especially for magnetised NSs. In particular, \citet{Dexheimer2017} showed that the magnetic field profile can be well approximated by a quadratic polynomial in the baryon chemical potential instead of an exponential function of matter density. They presented a realistic distribution for a poloidal magnetic field in the polar direction, for which the magnetic field distribution self-consistently satisfies the Einstein-Maxwell field equations.

There is no hydrogen burning or other nuclear fusion reactions taking place within the WD core, so we assume that the radial luminosity is constant $L_r=L$, where $L$ is the surface luminosity. The degenerate electrons in the WD core generally have a large mean free path due to the filled Fermi sea and therefore their high thermal conductivity leads to a uniform temperature throughout the region \citep{ST1983}. 
In the case of magnetised NSs, the thermal conduction can be suppressed along directions transverse to the magnetic field lines \citep{Hernquist85,PCY2007}. However, these changes in conduction rates are unlikely to affect the cooling process in B-WDs because the insulating region is non-degenerate and thermal conduction occurs only in the stellar interior \citep{Tremblay2015}. Furthermore, the average magnetic fields considered here are much weaker than those in magnetised neutron stars (NSs) and so we assume that the core is perfectly isothermal for B-WDs. \citet{Gupta2020} considered speculative cases with $dT/dr \neq 0$ below the interface to show that even non-magnetised WDs can have super-Chandrasekhar masses for sufficiently large luminosities. However, for the purpose of our work here, we ignore this possibility because there is no existing observational evidence to indicate that non-magnetised and/or non-rotating WDs can have super-Chandrasekhar masses.

\subsection{Nonmagnetic results}
First we explore some basic features of nonmagnetic WDs considering $B=(0,0)$.
We solve the set of differential equations (1) to (3) for model WDs with Runge-Kutta method, by providing the surface density, mass and surface temperature as the boundary conditions. For a given WD surface luminosity $L$ and its corresponding radius $R$, the surface temperature is obtained with the Stefan-Boltzmann law as $T_{\rm s} = (L/4\pi R^2 \sigma)^{1/4}$, where $\sigma$ is the Stefan-Boltzmann constant. We set the surface density to $\rho_s = 10^{-4}\ {\rm g\ cm^{-3}}$ as representative for all the cases considered here. Following \citet{Gupta2020}, the total mass is obtained by iteratively solving the equations (1) to (3) until the integrated mass starting from the WD surface and extending up to $10\, {\rm km}$ from the centre matches the mass that is obtained by solving the mass conservation equation (3) independently using the solution for the density profile. 

The left panel of Figure 1 shows the effect of luminosity and thereby the temperature gradient (see equation 2) on the mass--radius relation for non-magnetised WDs compared to Chandrasekhar's results \citep{Chandrasekhar1935}. 
We find that the Chandrasekhar mass limit is retained irrespective of the luminosity $10^{-4} \lesssim L/L_{\odot} \lesssim 10^{-2}$. However, increase in surface luminosity leads to progressively higher masses for the larger WDs. This is expected as larger $L$ translates to more thermal energy which results in higher ideal gas pressure, thereby allowing the WD to hold more mass. 
The right panel of Figure 1 shows the radial temperature profiles corresponding to the same luminosities for a $R=10000\, {\rm km}$ WD.
While the uniform $T$ within the isothermal core increases with a corresponding increase in $L$, the temperature drops rapidly within the thin non-degenerate surface layers. With the increase in $L$, the interface shifts inwards and the degenerate region shrinks in volume. 
As expected from equation (2), the temperature gradient near the surface increases with $L$.

Table 1 lists the central and interface properties for a range of WD radii $2000 \lesssim R/{\rm km} \lesssim 20000$ and luminosities $10^{-4} \lesssim L/L_{\odot} \lesssim 10^{-2}$. 
It can be seen that an increase in $L$ for a given $R$ leads to somewhat larger central densities $\rho_c$ but a more compact degenerate core (smaller interface radius $R_*$) and therefore an increased capacity to retain more mass. This effect tends to be more pronounced for larger radius WDs. 
In the case of smaller stars, the increase in central density due to larger effective thermal energy (or higher $L$) is not found to be significant such that the Chandrasekhar mass limit is always preserved. 
For the same $L$ but a larger $R$, the core temperatures are approximately similar but $\rho_c$ decreases substantially.

\section{Magnetic field effects}
In the previous section, we have described the formalism used to self-consistently obtain the properties of WDs by solving the structure equations for a given luminosity and radius and also obtained the solutions for nonmagnetised WDs. Here we discuss in detail the effects of magnetic field on the thermal properties as well as the mass--radius relations of B-WDs, for both the weak and strong magnetic field limits. Based on the strength of the magnetic field, we appropriately modify our treatment of the radiative opacity, magnetostatic pressure balance and EoS for the degenerate core and thence the interface location.

\begin{figure}
\includegraphics[width=\columnwidth]{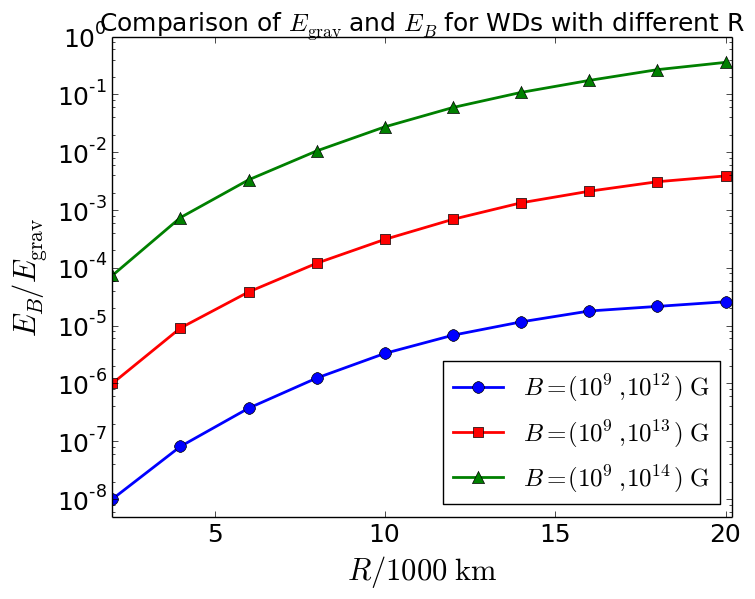} 
\caption{
Comparison between magnetic field energy and gravitational energy within the degenerate core used in order to determine the appropriate field regimes for B-WDs with different radii. The results here are shown for varying magnetic fields $B=(10^9,10^{12})\ {\rm G}$ (blue circles), $B=(10^9,10^{13})\ {\rm G}$ (red squares) and $B=(10^9,10^{14})\ {\rm G}$ (green triangles).
} 
\label{fig4} 
\end{figure}

We compare the energy density in the magnetic field with that of the gravitational field in order to distinguish between the weakly and strongly magnetised cases. We only evaluate the contribution to the respective energy densities from the degenerate core because the envelope is generally very thin with a high magnetic field and theoretically vanishing matter density for the cases that we consider here. 
Assuming a polytropic stellar structure model with an index $n=3$ (corresponding to
$\gamma=4/3$), the WD central density can be evaluated for a general solution of the Lane--Emden equation as $\rho_c = M/0.077 R^3 \approx 3.636\times10^7\ {\rm g\ cm^{-3}}(M/1.4M_{\odot})(R/10000\ {\rm km})^{-3}$. For a spherically symmetric WD geometry, the average gravitational energy density is $\rho_{\rm m,avg} \approx 3M/4\pi R^3 = 6.688\times10^5\ {\rm g\ cm^{-3}}(M/1.4M_{\odot})(R/10000\ {\rm km})^{-3}$, while the average magnetic field energy density is $\rho_{\rm B,avg}=44.232\ {\rm g\ cm^{-3}}(B_{\rm avg}/10^{12}\ {\rm G})^2$, where $B_{\rm avg}$ is obtained from the field profile (see equation 4) averaged over the matter density assuming $\rho_0 \approx 0.1\rho_c$. We consider the strong magnetic field limit to be valid provided that  $\zeta = \rho_{\rm B,avg}/\rho_{\rm m,avg} \gtrsim 0.01$ for a given WD mass and radius. 

Figure 2 shows the ratio between the magnetic field and gravitational energies within the isothermal degenerate B-WD core for three different magnetic field configurations and radius within $2000 \lesssim R/{\rm km} \lesssim 20000$. 
We find that the effect of surface field on the ratio $E_B/E_{\rm grav}$ is insignificant provided that the central field is considerably larger than the surface magnetic field. 
As shown in Figure 2, the gravitational energy dominates over the magnetic field energy for the entire range of WD radius and for central fields as strong as $\sim 10^{14}\, {\rm G}$. 
While assuming a spherically symmetric star is justified provided that $B \lesssim 10^{14}\, {\rm G}$, WDs with such strong magnetic fields can often have significantly larger masses as compared to their non-magnetised counterparts (see \citealt{SM2015}). This would then result in a higher gravitational energy density relative to magnetic energy density and consequently marginally smaller $E_B/E_{\rm grav}$ ratios than what we estimate for the $B=(10^9,10^{14})\, {\rm G}$ case here.
Therefore, the B-WD configurations considered in our study will always satisfy the structural stability criteria for both weak and strong magnetic field limits.

\subsection{Weak field limit}
For weaker magnetic fields with $\zeta \ll 1$, the radiative opacity can be effectively approximated with Kramers' formula, $\kappa = \kappa_0 \rho T^{-3.5}$, similar to the non-magnetised WDs discussed in Section 2. Similarly, the general relativistic effects can also be ignored while considering the magnetostatic pressure balance (see equation 1) because the incremental change to the inferred B-WD mass is not found to be significant \citep{Gupta2020}. While the EoS for the B-WD core matter corresponds to that of a non-relativistic degenerate gas, the surface layer EoS is given by non-degenerate ideal gas. In order to evaluate the interface location, the respective pressures can be equated on both sides to obtain \citep{ST1983}
\be
\rho_* \approx (2.4\times10^{-8}\ {\rm g\ cm^{-3}}\ K^{-3/2})\mu_e T_*^{3/2}, \hspace{1cm}
\label{int_weakB}
\ee
where $\mu_e \approx 2$ is the mean molecular weight per electron, $\rho_*$ and $T_*$ are the density and temperature respectively for the interface between the degenerate core and the non-degenerate envelope.

\subsection{Strong field limit}
Strong magnetic fields can modify the radiative opacity for photon diffusion as well as the EoS of the matter therein. For sufficiently large magnetic fields, the variation of radiative opacity with $B$ can be modelled similarly to neutron stars as $\kappa = \kappa_B \approx 5.5\times10^{31}\rho T^{-1.5} B^{-2} {\rm cm^{2}\ g^{-1}}$ \citep{PY2001,VP2001}. The magnetic field dependent Potekhin's opacity is generally used instead of the Kramers' opacity if $B/10^{12}\ {\rm G} \geq T/10^6\ {\rm K}$ and if radiation dominates over convection which is valid for the strong $B$ cases that we consider here. The general relativistic effects on the inferred mass--radius relation of B-WDs have already been investigated in detail for various magnetic field and rotational configurations \citep{DM2014,SM2015}. The effect of poloidally as well as toroidally dominated field configurations is seen as super-Chandrasekhar WDs with generally non-spherical shapes that depend on the magnetic field and geometry.

\begin{figure}
\includegraphics[width=\columnwidth]{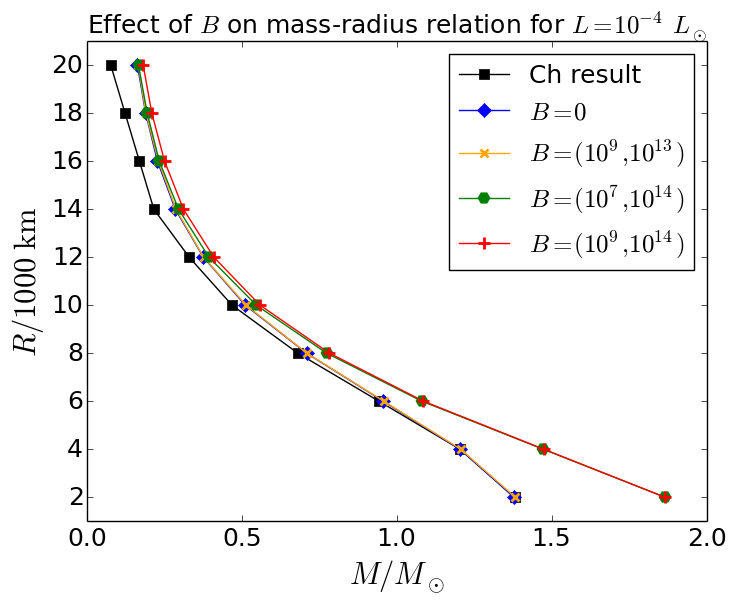}
\caption{The effect of magnetic field on the mass--radius relation of B-WDs is shown for $B=(0,0)$ (blue diamonds), $B=(10^9,10^{13})\ {\rm G}$ (orange crosses), $B=(10^7,10^{14})\ {\rm G}$ (green circles) and $B=(10^9,10^{14})\ {\rm G}$ (red pluses),
along with the Chandrasekhar result (black squares). The luminosity is set as $10^{-4}\ L_{\odot}$ for these results.
} 
\label{fig2} 
\end{figure}

\begin{table*}
\caption{The effect of magnetic field $B$ on the mass--radius relation for luminosity $L = 10^{-4}\ L_{\odot}$. The surface field $B_{\rm s}$ and fiducial field $B_0$ are defined as in equation \ref{Bprof}.
}
\label{Table2}
\bgroup
\def\arraystretch{1.0}
\begin{tabular}{| c | c | c | c | c | c | c |}
\hline
\hline
\centering
$R/1000\ {\rm km}$ & $(B_s,B_0)/{\rm G}$ & $T_c/10^6 \ {\rm K}$ & $\rho_c/10^6 \ {\rm g\ cm^{-3}}$ & $\rho_*/10^6 \ {\rm g\ cm^{-3}}$ & $R_*/1000\ {\rm km}$ & $M/M_{\odot}$ \\ \hline \hline
2.0 & $(10^7,10^{13})$ & 4.51 & 2234 & 0.0007 & 1.997 & 1.382 \\ 
    & $(10^9,10^{13})$ & 4.58 & 2240 & 0.0007 & 1.997 & 1.382 \\
    & $(10^7,10^{14})$ & 4.34 & 2919 & 0.0005 & 1.998 & 1.865 \\ 
    & $(10^9,10^{14})$ & 4.26 & 2927 & 0.0006 & 1.996 & 1.865 \\\hline
6.0 & $(10^7,10^{13})$ & 5.04 & 24.94 & 0.0007 & 5.958 & 0.958 \\ 
    & $(10^9,10^{13})$ & 4.96 & 24.71 & 0.0006 & 5.958 & 0.958 \\
    & $(10^7,10^{14})$ & 4.89 & 26.87 & 0.0005 & 5.967 & 1.081 \\ 
    & $(10^9,10^{14})$ & 4.83 & 26.58 & 0.0005 & 5.952 & 1.086 \\\hline
12.0 & $(10^7,10^{13})$ & 6.53 & 0.837 & 0.0012 & 11.446 & 0.374 \\ 
    & $(10^9,10^{13})$ & 6.60 & 0.856 & 0.0011 & 11.412 & 0.377 \\
    & $(10^7,10^{14})$ & 6.33 & 0.893 & 0.0011 & 11.468 & 0.394 \\ 
    & $(10^9,10^{14})$ & 6.52 & 0.962 & 0.0012 & 11.253 & 0.409 \\ \hline 
20.0 & $(10^7,10^{13})$ & 8.22 & 0.070 & 0.0015 & 16.261 & 0.164 \\ 
    & $(10^9,10^{13})$ & 8.41 & 0.064 & 0.0014 & 16.275 & 0.166 \\
    & $(10^7,10^{14})$ & 8.39 & 0.099 & 0.0013 & 16.045 & 0.167 \\ 
    & $(10^9,10^{14})$ & 8.58 & 0.126 & 0.0015 & 15.852 & 0.182 \\ \hline \hline
\end{tabular}
\egroup
\end{table*}

Here we explore the general relativistic effects on the B-WD with the Tolman-Oppenheimer-Volkoff equation \citep{OV1939}
\be
\frac{d\rho}{dr} = -\frac{G[\rho + \rho_B + (P+P_B)/c^2][m(r)+ 4\pi r^3 (P+P_B)/c^2]}{[r^2 - 2Gm(r)r/c^2](dP/d\rho + dP_B/d\rho)},
\label{tov}
\ee
where $P = P_{\rm deg} + P_{\rm ig}$ is the sum of the electron degeneracy pressure and ideal gas pressure. The remaining terms are defined similarly to equation (1). However, for simplicity, here we assume the WDs to be approximately spherical in shape and neglect the magnetic tension term that can lead to potentially anisotropic pressure component. Hence equation (\ref{tov}) is valid only approximately. Nevertheless, recent works \citep{SM2015,KM2019} showed that, for toroidally dominated magnetic fields, a B-WD does not deviate much from a sphere and the natural existence of toroidally dominated fields was indeed confirmed by \citet{QY2018} with numerical simulations. 

In the presence of strong fields $\zeta \gtrsim 0.01$, quantum mechanical effects turn out to be important and equation (5) is not strictly valid while obtaining the interface radius \citep{HPY2007}. After including the Landau quantisation effects, the degenerate core EoS depends on the magnetic field whereas the non-degenerate envelope EoS remains unaffected \citep{VP2001}. The electron pressure can then be equated on both sides to give
\be
\rho_{*}(B_{*}) = (1.482\times10^{-12}\ {\rm g\ cm^{-3}\ K^{-1/2}\ G^{-1}}) T_{*}^{1/2} B_{*},
\label{int_strongB}
\ee
where $B_*(\rho_*)$ is the magnetic field at the interface radius $r=r_*$. It has already been shown \citep{DM2012,DM2013} that if $B > B_{\rm c} \approx 4.414\times10^{13}\ {\rm G}$, the electron Larmor radius becomes comparable to the Compton wavelength and the electron degenerate matter EoS has to be adequately modified. This can potentially yield super-Chandrasekhar WDs with mass-limit $M \approx 2.58\ M_{\odot}$.

\begin{figure*} 
\includegraphics[width=0.49\textwidth]{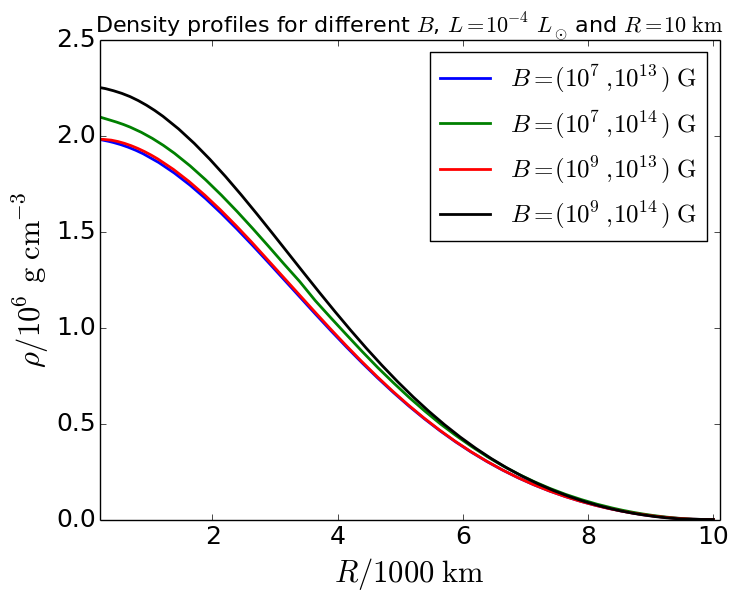} 
\includegraphics[width=0.49\textwidth]{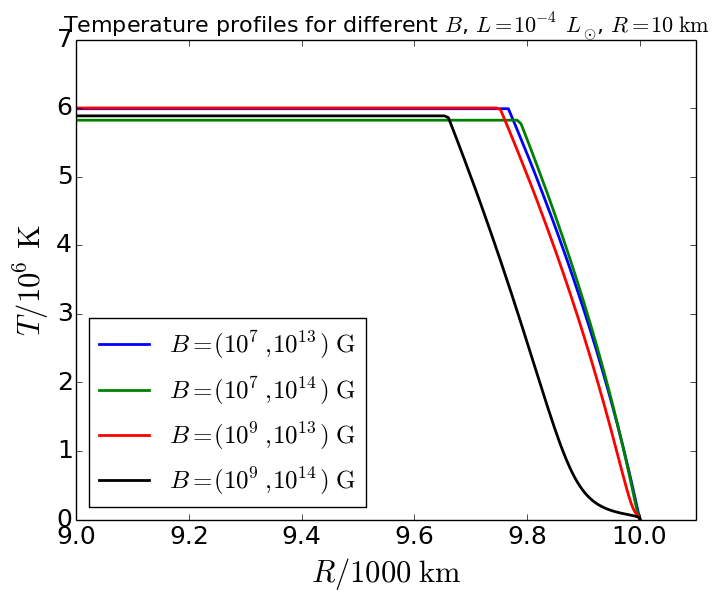} 
\caption{
{\bf Left panel:} The variation of density as a function of radius is shown for WDs with varying magnetic fields $B=(10^7,10^{13})\ {\rm G}$ (blue), $B=(10^7,10^{14})\ {\rm G}$ (green), $B=(10^9,10^{13})\ {\rm G}$ (red) and $B=(10^9,10^{14})\ {\rm G}$ (black).
{\bf Right panel:} The variation of temperature as a function of radius is shown for the same field configurations as in the left panel. The luminosity is set as $10^{-4}\ L_{\odot}$ for all these results. 
} 
\label{fig3}
\end{figure*}

\subsection{Results}
Here we discuss the results related to the mass--radius relation and WD structure properties for both weak and strong magnetic field cases. Figure 3 shows the effect of magnetic field on the mass--radius relation for B-WDs with surface luminosity $L=10^{-4}\, L_{\odot}$ and compares them to the non-magnetic Chandrasekhar results. 
We find that the magnetic field affects the mass--radius relation in a manner that is analogous to increasing $L$ (see Figure 1), by shifting the curve towards higher masses for WDs with larger radii. 
While $B_{\rm s}$ does not have any appreciable effect on the inferred mass, the magnitude of $B_0$ is vital in deciding the shape of the mass--radius curve. 
The mass--radius curves for $B_0 \lesssim 10^{13}\, {\rm G}$ practically overlap with each other and retain the Chandrasekhar mass limit. However, for strong central fields with $B_0 \sim 10^{14}\, {\rm G}$, we obtain super-Chandrasekhar WDs with masses as large as $\sim 1.9\, M_{\odot}$.

Table 2 lists the central and interface properties for different magnetic fields and B-WD radii for fixed luminosity $L=10^{-4}\, L_{\odot}$. While the interface location $R_*$ and the interface density $\rho_*$ do not vary much with $B$, the core density $\rho_c$ and therefore the total mass $M$ increase significantly with an increase in the magnetic field. 
The central density for stronger fields is expected to be larger in order to compensate for the additional magnetic pressure, as the relative contribution of magnetic field to the total pressure is larger than that to the total density. With increase in stellar radius for a given $B$, $\rho_c$ reduces substantially whereas $R_*$ increases, resulting in smaller average mass density within the isothermal degenerate core. 
The left panel of Figure 4 shows the radial variation of the matter density for $R=10000\, {\rm km}$ and different magnetic field profiles, while the right panel shows the temperature profiles for the same cases. 
As listed in Table 2, the interface densities $\rho_*$ tend to be roughly similar but the central density $\rho_c$ increases with the effective magnetic field. 
Although the core temperature $T_c$ (primarily determined by $L$) is unchanged irrespective of the magnetic field, the radial temperature gradient $dT/dr$ within the surface layers for $B=(10^9,10^{14})\, {\rm G}$ is approximately half as that for the other weaker field cases.

\section{Luminosity suppression, cooling and field decay in B-WDs}

\begin{table*}
\caption{The effects of magnetic fields on the B-WD luminosity in order for the magnetised mass--radius relation to match with the non-magnetised relation. The initial field at $t=0$ is kept fixed at $B=(10^9,10^{14})\ {\rm G}$ for all the radii listed here. The topmost entry for each radius is for the initial time $t=0$ whereas the bottom two entries list the corresponding parameters for $t=\tau=10\ {\rm Gyr}$ after including the cooling rate and magnetic field decay over time. 
While we evaluate the magnetic fields assuming that Ohmic dissipation is the dominant process for the top entries of $\tau=10\ {\rm Gyr}$, for the bottom entries, we assume that Hall drift is the primary process until the field parameter $B_0$ decays to $\sim 10^{12}\ {\rm G}$, below which Ohmic dissipation dominates.
}
\label{Table3}
\bgroup
\def\arraystretch{1.0}
\begin{tabular}{| c | c | c | c | c | c | c | c |}
\hline
\hline
\centering
$R/1000\ {\rm km}$ & $t/{\rm Gyr}$ & $t_{\rm HO}/\tau$ 
& $B_{\rm s}/{\rm G}$ & $B_0/{\rm G}$ & $L/L_{\odot}$ & $M_{B=0}/M_{\odot}$ & $M/M_{\odot}$ \\ \hline \hline
2.0 & 0 & & $10^9$ & $10^{14}$ & $10^{-16}$ & 1.378 & 1.865 \\ \noalign{\smallskip} \cline{2-8} \noalign{\smallskip}
    & 10 & 0 & $4.58\times10^8$ & $4.58\times10^{13}$ & $10^{-16}$ & 1.377 & 1.478 \\
    & & 1 & & $5.83\times10^{13}$ & $10^{-16}$ & & 1.542 \\ \hline
4.0 & 0 & & $10^9$ & $10^{14}$ & $10^{-16}$ & 1.204 & 1.470 \\ \noalign{\smallskip} \cline{2-8} \noalign{\smallskip} 
    & 10 & 0 & $2.78\times10^8$ & $2.78\times10^{13}$ & $10^{-12}$ & 1.201 & 1.218 \\ 
    & & 0.224 & & $3.71\times10^{11}$ & $3\times10^{-7}$ & & 1.201 \\ \hline
6.0 & 0 & & $10^9$ & $10^{14}$ & $10^{-16}$ & 0.956 & 1.074 \\ \noalign{\smallskip} \cline{2-8} \noalign{\smallskip} 
    & 10 & 0 & $1.65\times10^8$ & $1.65\times10^{13}$ & $10^{-8}$ &
    0.951 & 0.951 \\
    & & $6.89\times10^{-2}$ & & $1.86\times10^{11}$ & $2\times10^{-6}$ & & 0.951 \\ \hline
8.0 & 0 & & $10^9$ & $10^{14}$ & $10^{-12}$ & 0.709 & 0.762 \\ \noalign{\smallskip} \cline{2-8} \noalign{\smallskip}
    & 10 & 0 & $9.86\times10^7$ & $9.86\times10^{12}$ & $2\times10^{-6}$ & 0.699 & 0.699 \\
    & & $2.28\times10^{-2}$ & & $1.04\times10^{11}$ & $7\times10^{-6}$ & & 0.699 \\ \hline
10.0 & 0 & & $10^9$ & $10^{14}$ & $10^{-12}$ & 0.512 & 0.527 \\ \noalign{\smallskip} \cline{2-8} \noalign{\smallskip}
    & 10 & 0 & $6.02\times10^7$ & $6.02\times10^{12}$ & $8\times10^{-6}$ & 0.496 & 0.496 \\ 
    & & $9.72\times10^{-3}$ & & $6.18\times10^{10}$ & $10^{-5}$ & & 0.496 \\ \hline
12.0 & 0 & & $10^9$ & $10^{14}$ & $7\times10^{-8}$ & 0.376 & 0.376 \\ \noalign{\smallskip} \cline{2-8} \noalign{\smallskip} 
    & 10 & 0 & $3.69\times10^7$ & $3.69\times10^{12}$ & $10^{-5}$ & 0.354 & 0.354 \\
    & & $4.88\times10^{-3}$ & & $3.75\times10^{10}$ & $10^{-5}$ & & 0.354 \\ \hline
14.0 & 0 & & $10^9$ & $10^{14}$ & $2\times10^{-6}$ & 0.286 & 0.286 \\ \noalign{\smallskip} \cline{2-8} \noalign{\smallskip}
    & 10 & 0 & $3.06\times10^7$ & $3.06\times10^{12}$ & $10^{-5}$ & 0.262 & 0.262 \\ 
    & & $2.22\times10^{-3}$ & & $3.08\times10^{10}$ & $10^{-5}$ & & 0.262 \\ \hline
16.0 & 0 & & $10^9$ & $10^{14}$ & $4\times10^{-6}$ & 0.228 & 0.228 \\ \noalign{\smallskip} \cline{2-8} \noalign{\smallskip}
    & 10 & 0 & $2.00\times10^7$ & $2.00\times10^{12}$ & $10^{-5}$ & 0.204 & 0.204 \\ 
    & & $1.34\times10^{-3}$ & & $2.01\times10^{10}$ & $10^{-5}$ & & 0.204 \\ \hline
18.0 & 0 & & $10^9$ & $10^{14}$ & $5\times10^{-6}$ & 0.190 & 0.190 \\ \noalign{\smallskip} \cline{2-8} \noalign{\smallskip}
    & 10 & 0 & $1.87\times10^7$ & $1.87\times10^{12}$ & $10^{-5}$ & 0.165 & 0.165 \\ 
    & & $7.75\times10^{-4}$ & & $1.88\times10^{10}$ & $10^{-5}$ & & 0.165 \\ \hline
20.0 & 0 & & $10^9$ & $10^{14}$ & $7\times10^{-6}$ & 0.164 & 0.164 \\ \noalign{\smallskip} \cline{2-8} \noalign{\smallskip}
    & 10 & 0 & $2.97\times10^7$ & $2.97\times10^{12}$ & $10^{-5}$ & 0.138 & 0.138 \\ 
    & & $3.70\times10^{-4}$ & & $2.98\times10^{10}$ & $10^{-5}$ & & 0.138 \\ \hline \hline
\end{tabular}
\egroup
\end{table*}

We study the effects of strong magnetic fields on the observed luminosities for magnetised WDs, i.e. B-WDs. In particular, we compute the luminosities that are expected for B-WDs in order to obtain mass--radius relations which are similar to those for non-magnetised WDs. We then discuss the cooling process of these B-WDs and also the decay of large magnetic fields by various dissipative processes including Ohmic dissipation, ambipolar diffusion and Hall drift that can occur within the typical WD age evolution.

\subsection{Suppression of luminosity}
\citet{MBh2018} investigated the variation of surface luminosity as the magnetic field increases, particularly for fixed interface radii and/or temperatures. The motivation for fixing interface parameters between non-magnetised and magnetised WDs was to better constrain the individual components (thermal, gravitational and magnetic) of the conserved total energy for these stars. Here, we relax the assumptions of fixed interface parameters between non-magnetised and magnetised WDs. Nevertheless, in order to ensure structural stability for a B-WD, an increase in magnetic energy density has to be compensated by a corresponding decrease in the thermal energy and hence the luminosity, provided the gravitational energy is not affected significantly. This effect is especially prominent for B-WDs with larger radii where the magnetic, thermal and gravitational energies are comparable with each other. 

Table 3 lists the initial luminosities (see rows with time $t=0$) corresponding to field $B=(10^9,10^{14})\, {\rm G}$ that yield masses closest to those obtained for non-magnetised WDs with radii $2000 \leq R/{\rm km} \leq 20000$. While a slight decrease in the luminosity (within the observable range) for $R \gtrsim 12000\, {\rm km}$ WDs leads to masses that are similar to those of their non-magnetic counterparts, the smaller radii B-WDs require a substantial drop in their luminosity (well outside the observable range) and still do not achieve masses that are similar to the non-magnetised WDs. This is expected because the thermal pressure is sub-dominant in case of WDs with small radii and therefore a decrease in the thermal energy (or luminosity) does not significantly affect the total mass. As a result, for stars with $2000 \leq R/{\rm km} \leq 10000$, even if the luminosity decreases substantially $10^{-16} \leq L/L_{\odot} \leq 10^{-12}$, the resulting mass of the B-WD remains larger than its non-magnetic counterpart. This leads to an extended branch in the mass--radius relation.

\subsection{Magnetic field decay}
In a strongly magnetised NS, ambipolar diffusion and beta decays can cause the magnetic energy release that is observed from magnetars. However, the magnetic fields inside an electron degenerate WD generally undergo decay by Ohmic dissipation and Hall drift processes with timescales that are given by \citet{HK1998} and \citet{Cumming2002} as
\bea
t_{\rm Ohm} = (7\times10^{10}\ {\rm yr})\, \rho_{c,6}^{1/3} R_{4}^{1/2} (\rho_{\rm avg}/\rho_{\rm c}),\\
t_{\rm Hall} = (5\times10^{10}\, {\rm yr})\ l_8^2 B_{0,14}^{-1} T_{\rm c,7}^{2} \rho_{\rm c,10},
\label{tOhm_tHall} 
\eea
where $\rho_{\rm c,n} = \rho_c/10^n\, {\rm g\, cm^{-3}}$, $R_4 = R/10^4\, {\rm km}$, $T_{c,7}=T_c/10^7\, {\rm K}$, $B_{0,14}=B_0/10^{14}\, {\rm G}$ and $l = l_8 \times 10^8\, {\rm cm}$ is a characteristic length scale of the flux loops through the outer core of the WD. 
In the case of isolated and cool WDs, theoretical calculations indicate that if the magnetic field is not very strong like that for the B-WD center, it typically decays due to Ohmic dissipation by a factor of two in $10\, {\rm Gyr}$ \citep{Fontaine1973,Wendell1987}.
\citet{Cumming2002} estimated a lowest order decay time of $8 \leq t_{\rm Ohm}/{\rm Gyr} \leq 12$ for dipole fields and $4 \leq t_{\rm Ohm}/{\rm Gyr} \leq 6$ for quadrupole fields in the context of accreting WDs. 

Ohmic decay is characterised by the induction equation, $\partial \mathbf{B}/\partial t = -\nabla \times (\eta\nabla \times \mathbf{B})$, where $\eta=c^2/4\pi \sigma$ is the magnetic diffusivity and $\sigma$ is the electrical conductivity. The field decay timescale can then be written as $t_{\rm Ohm} \approx 4\pi \sigma L^2/c^2$, where $L$ the length scale over which the field varies \citep{Cumming2002}. The electrical conductivity is set by the collisions between the electrons and ions (see, e.g., \citealt{YU1980,Itoh1983,Schatz1999}). Unlike within the degenerate core, the electrical conductivity is dependent on the temperature in the surface layers where the electrons are non-degenerate. As the B-WD mass increases, the increase in conductivity is offset by the decreasing radius.

Ohmic decay is the dominant field dissipation process for $B \lesssim 10^{12}\ {\rm G}$, while for $10^{12} \leq B/{\rm G} \leq 10^{14}$ the decay occurs via Hall drift and for $B \gtrsim 10^{14}\ {\rm G}$, the principal decay mechanism is likely to be ambipolar diffusion \citep{HK1998}. We assume that Hall drift dominates within the B-WD degenerate core with $B_0 \approx 10^{14}\, {\rm G}$ while Ohmic dissipation is the main decay mechanism for surface fields $B_{\rm s} \approx 10^9\, {\rm G}$. 
Although equations (8) and (9) are primarily used to model the field decay over time for strongly magnetised NSs, here we appropriately adopt them for typical B-WD structure properties.

The magnetic field decay in magnetars with surface fields between $10^{14}$ and $10^{16}\, {\rm G}$ was studied, using an appropriate cooling model by \citet{HK1998} and by solving the decay equation
\be
\frac{{\rm d}B}{{\rm d}t} = -B\left(\frac{1}{t_{\rm Ohm}} + \frac{1}{t_{\rm Amb}} + \frac{1}{t_{\rm Hall}}\right),
\ee
where $t_{\rm Amb}$ denotes the ambipolar diffusion time scale. Because we consider strongly magnetised WDs with central fields of about $10^{14}\, {\rm G}$, comparable to the surface fields of magnetars, we assume that the magnetic fields in magnetars and B-WDs both undergo similar decay mechanisms. \citet{Muslimov1995} showed that Hall drift is not expected to be a direct cause of magnetic field decay in WDs because it conserves the total magnetic energy. However, in the presence of magnetic turbulence, Hall drift can twist the field lines and thence enhance Ohmic dissipation \citep{GR1992}. To model and compare the relative contributions from these different processes, we consider two separate cases: (a) when only Ohmic dissipation occurs for both the surface and central magnetic fields, (b) while $B_{\rm s}$ continues to evolve over $t_{\rm Ohm}$, Hall drift determines the $B_0$ evolution until the central field drops to about $10^{12}\, {\rm G}$, below which Ohmic dissipation sets in.

\begin{figure}
\includegraphics[width=\columnwidth]{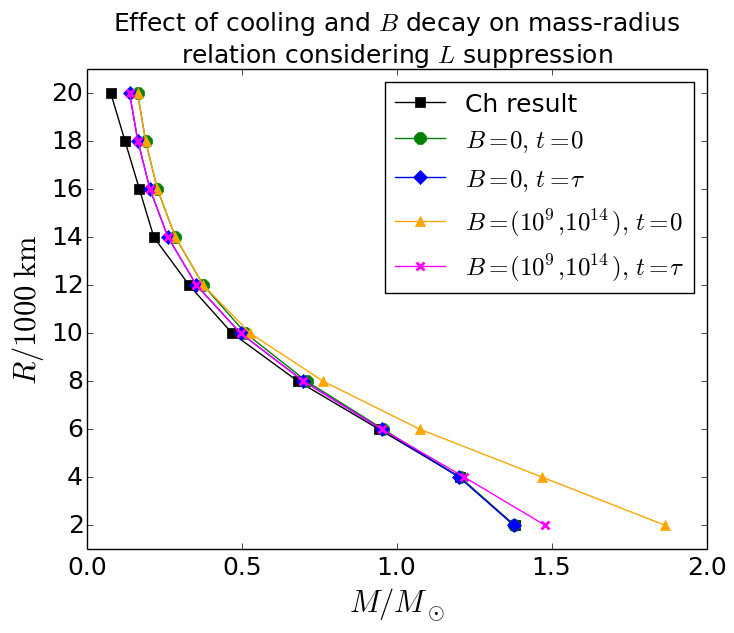}
\caption{
The effect of magnetic field on the WD luminosity set to match with the non-magnetised mass--radius relation. The results are shown for $B=(0,0)$ at initial time $t=0$ (green circles), $B=(0,0)$ at $t=10\ {\rm Gyr}$ (blue diamonds), $B=(10^9,10^{14})\, {\rm G}$ at $t=0$ (orange triangles) and $B=(10^9,10^{14})\, {\rm G}$ at $t=10\ {\rm Gyr}$ (magenta crosses). The time evolution results are obtained after including B-WD cooling evolution and field dissipation processes. For the magenta curve, we only show the Ohmic dissipation results because the results with Hall drift decay practically overlap with them (see Table 3 for the specific luminosities).
} 
\label{fig5} 
\end{figure}

\subsection{B-WD cooling}
The cooling evolution of non-magnetised WDs has been investigated in detail with theoretical attempts to model the degenerate core as the primary energy source \citep{Mestel1952,MR1967}. The thermal energy is radiated away gradually over time in the observed luminosity from the surface layers as the star evolves. \citet{TY1996} computed the cooling curves for low-mass WDs, starting as luminous stars, until their crystallisation stage after about $10\, {\rm Gyr}$. Because most of the electrons occupy the lowest energy states in a degenerate gas, the thermal energy of ions is the only significant energy source that can be radiated. The degenerate electrons in the interior of B-WDs have large mean free path which leads to high thermal conductivity and therefore uniform temperature. The isothermal interior is covered by the non-degenerate surface layers which transport the energy flux outward by photon diffusion.

The interface location between the non-degenerate surface layers and the degenerate interior is obtained by equating the pressure from both sides, as done in equations (\ref{int_weakB}) and (\ref{int_strongB}) for the weak field and strong field cases, respectively. The luminosity is related to the interface temperature as $L \propto T^{7/2}/\kappa$ \citep{ST1983}, where the opacity $\kappa$ depends on the field strength (see Section 3). The uniform interior temperature is then computed for a given luminosity, composition and mass of the B-WD. 
The rate at which the thermal energy of ions can be transported to the surface and thence to be radiated depends on the specific heat \citep{ST1983}, and is given by
\be
L = -\frac{d}{dt}\int c_{\rm v} dT = (2\times10^6\ {\rm erg/s)}\, \frac{Am_{\mu}}{M_{\odot}}\left(\frac{T}{K}\right)^{7/2},
\label{L_T_relation}
\ee
where $c_v \approx 3k_B/2$ is the specific heat at constant volume, $m_{\mu}$ is the proton mass and $A$ is the atomic weight. Given an initial luminosity and temperature $T_0$ at time $t_0$, the final temperature $T$ at time $t$ after cooling is obtained by directly integrating equation (\ref{L_T_relation}) as
\be
(T/{\rm K})^{-5/2} - (T_0/{\rm K})^{-5/2} = 2.406\times10^{-34}\, \tau/{\rm s},
\ee
where $\tau = t-t_0$ is the WD age.

Although convection can aid faster cooling with a more efficient energy transport, its effect has been shown to be insignificant in a first-order approximation \citep{LVH1975,FVH1976}. This is due to the fact that convection only influences the cooling time once the base of the convection zone reaches the degenerate thermal energy reservoir and couples the surface to the reservoir. However, this is the case only for significantly lower surface temperatures than we consider here. \citet{Tremblay2015} recently showed that convective energy transfer can be significantly impeded once the magnetic pressure dominates over the thermal pressure. It is important to note that, for simplicity in the calculations here, we have assumed self-similarity of the cooling process over the entire evolution of the WD. However, a more detailed calculation of non-magnetised WD cooling has shown that this might not strictly be the case \citep{Hansen1999}.

\begin{figure*} 
\includegraphics[width=0.48\textwidth]{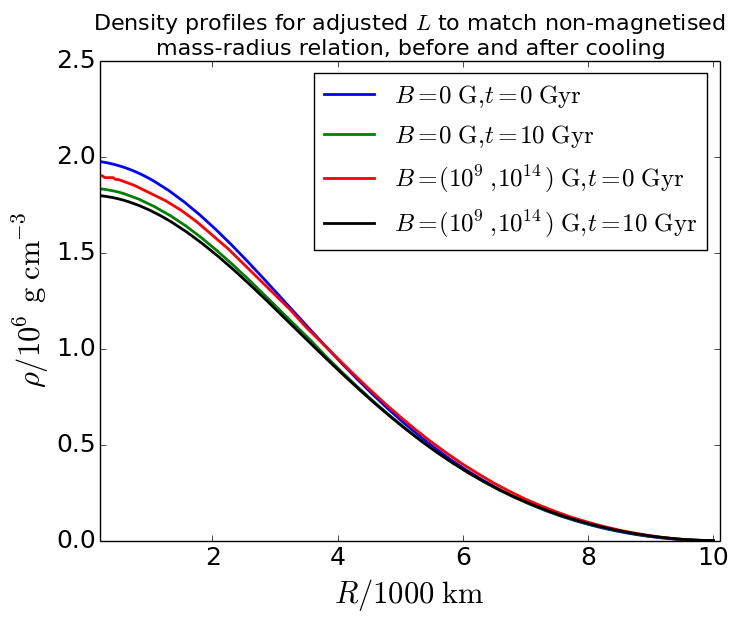} 
\includegraphics[width=0.50\textwidth]{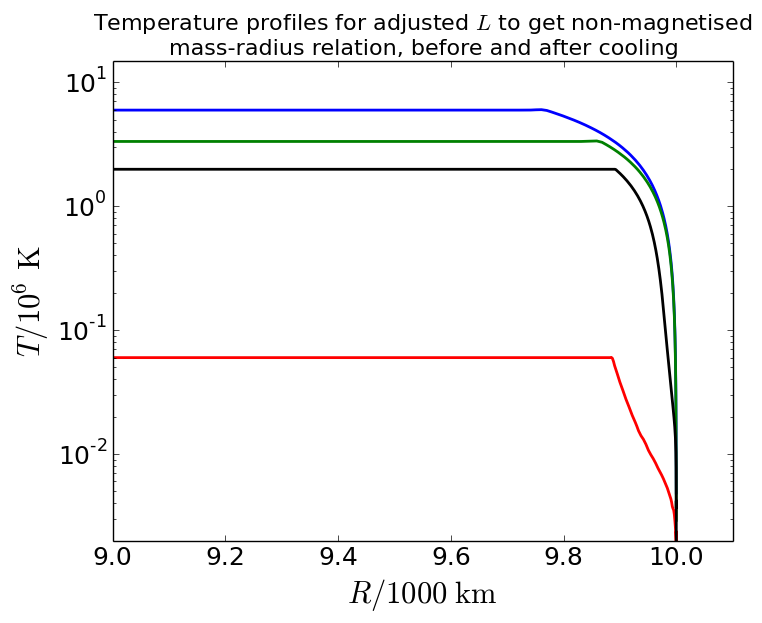} 
\caption{{\bf Left panel:}
The variation of density is shown as a function of radius for the same cases as in Figure 5 and $R=10000\ {\rm km}$.
{\bf Right panel:} The variation of temperature with radius are shown for the same cases as Figure 5 and $R=10000\ {\rm km}$. The corresponding luminosities are listed in Table 3.
} 
\label{fig6}
\end{figure*}

In the case of magnetised WDs, the relevant parameter affecting the state of the ionic core and therefore its thermodynamic properties is $b = \omega_B/\omega_p$, where $\omega_B = ZeB/\tilde{M}c$ and $\omega_{\rm p} = \sqrt{4\pi Z^2 e^2 n/\tilde{M}}$ are the ion cyclotron and plasma frequencies, respectively, $Z$ is the atomic number, $\tilde{M}$ is the mass of the nuclei, $n$ is the ion number density and $e$ the electronic charge. 
The effect of a magnetic field on the ionic core is expected to be strong when the cyclotron frequency is comparable to or larger than the lattice Debye frequency for which $b \gtrsim 1$. \citet{Baiko2009} studied the effect of magnetic fields on ionic lattices and showed that there is an appreciable change to the specific heat only when $b \gg 1$ except when $T \ll \theta_{\rm D}$ (Debye temperature). For the magnetic field configurations that we consider here $b \leq 1$ and the interface temperature is comparable to $\theta_{\rm D}$. So, we assume a specific heat model that is the same as that for non-magnetised WDs despite the  magnetic fields. The effect of magnetic fields on the phonon spectrum of ions in conventional systems has been studied and found to be generally weak \citep{Holz1972}.

\subsection{Results}
Table 3 lists the luminosities and masses corresponding to WD radii $2000 \lesssim R/{\rm km} \lesssim 20000$ for initial $B=(10^9,10^{14})\, {\rm G}$ at time $t=0$ and $t=10\, {\rm Gyr}$.
The top entry for each radius is for time $t=0$ and the bottom two entries list the corresponding quantities after cooling for $t=10\, {\rm Gyr}$, also accounting for  simultaneous magnetic field dissipation.  
We consider two possibilities for the field decay: (a) Ohmic dissipation dominates for the entire evolution, (b) Hall drift is the primary field decay mechanism until $B_0$ drops below $10^{12}\, {\rm G}$, when Ohmic dissipation sets in. 
Based on the B-WD structure parameters (see equations 8 and 9), we find that the fraction of time $t_{\rm HO}/\tau$, when the Hall drift dominates, falls significantly with  increasing stellar radius. Consequently, the magnetic field decays considerably more because the faster Ohmic dissipation process turns out to be critical for much of the cooling evolution. 
As a result of the field decay and simultaneous cooling over $10\, {\rm Gyr}$, the luminosity adjusted limiting B-WD masses are found to be substantially lower than their $t=0$ counterparts -- particularly for the smaller radius stars. We find that the mass--radius relations practically merge for $R \gtrsim 6000\, {\rm km}$ WDs. 
Furthermore, the inferred luminosities are also much less suppressed for the intermediate radii WDs with $6000 \lesssim R/{\rm km} \lesssim 12000$. Although the limiting masses for small B-WDs (with $R \approx 2000\, {\rm km}$) drop to about $1.5\ M_{\odot}$ compared to $1.9\ M_{\odot}$ without evolution, the majority of these systems still remain practically hidden throughout their cooling evolution because of their strong fields and correspondingly low $L$, outside the observable range.

Figure 5 shows the effect of the evolution of B-WDs on their mass--radius relations including both the magnetic field decay and thermal cooling effects. The luminosities are varied with the magnetic fields such that the B-WD masses can match those obtained for the non-magnetised WDs. We do not show the results for the Hall drift decay because they essentially overlap with the Ohmic dissipation results. 
For the $B=(0,0)$ case, we find that the mass--radius relation is shifted slightly more towards the Chandrasekhar result as a result of the cooling evolution and the mass limit remains unchanged. In case of $B=(10^9,10^{14})\, {\rm G}$, even though the limiting mass $\sim 1.9\, M_{\odot}$ at small radius turns out to be much larger than the Chandrasekhar limit of $1.4\, M_{\odot}$ (also see Table 3), we find that it is lowered considerably to $\sim 1.5\, M_{\odot}$ primarily as a result of magnetic field decay and also thermal cooling over $t=10\, {\rm Gyr}$. 

The left panel of Figure 6 shows the radial variation of the matter density for the same cases as shown in Figure 5 and $R=10000\, {\rm km}$, while the right panel shows the temperature profiles for the corresponding cases. We find that the matter density at the core is slightly suppressed in the presence of strong $B$ and also as a result of the evolutionary processes. As the total stellar energy is conserved at $t=0$, an increase in the magnetic energy has to be compensated by a similar decrease in the gravitational energy and hence the central density $\rho_{\rm c}$. 
Once the field decays and $L$ declines due to cooling, the central density adjusts itself to be slightly lower in order to balance the loss of magnetic and thermal energies with time. 
From the right panel of Figure 6, we see that the temperature $T_{\rm c}$ within the degenerate core is smaller for larger fields. This is expected as the luminosity needs to be suppressed more in order to have the total stellar energy fixed. However, owing to the appreciable field decay over $t=10\, {\rm Gyr}$, the luminosity increases by more than an order of magnitude. 
As the degenerate core volume is greater for magnetised WDs with a larger $R_*$, the temperature gradient $dT/dr$ within the envelope turns out to be considerably smaller for B-WDs than their non-magnetised counterparts.

\section{STARS results}
In this section, we describe the grid of numerical models that we have produced and analysed in order to investigate, compare and validate the analytical results described in Sections 2, 3 and 4. In Section 2, we have described the considerations made to self-consistently obtain the structural properties of magnetised WDs. These were used to produce a set of non-magnetic results to validate our results in the context of the existing literature. In Section 3, we have described the effects of the magnetic field on the mass--radius relations as well as the thermal properties of B-WDs. We have also described in detail the differences in our methodology and results between the weak and strong field cases. In Section~4, we have described the effects of strong magnetic fields on the observed luminosities of magnetised WDs. In particular, we have described the phenomenon of luminosity suppression, cooling and field decay in B-WDs.  

\subsection{Implementation and method}
In order to investigate the analytical prescription described in the previous sections, here we explore a set of numerical stellar evolution models using a modified version of the \textsc{stars} stellar evolution code \citep{EGG1971}.
The EoS solving subroutine \textsc{statef.f} is modified appropriately to include the prescriptions of \cite{Gupta2020} (also see \citealt{MBh2018}, who initiated this venture). This involves computing $B$ at each calculation shell and then computing the magnetic contribution to the pressure and density. For large fields such that $B/10^{12}\, {\rm G} \geq T/10^{6}\, {\rm K}$, the opacity is expected to be dominated by the field dependent Potekhin's opacity rather than the usual Kramers' opacity. Hence, alongside our usual tabulated opacity $\kappa_{\rm tab}$, which includes both OPAL opacities for the envelope \citep{Iglesias1996} and electron conduction \citep{Itoh1983}, we evaluate Potekhin's opacity $\kappa_B$ as discussed in Section 3.
The overall opacity is then computed by summing the tabulated opacity and the Potekhin opacity in inverse as $1/\kappa_{\rm tot} = 1/\kappa_{\rm tab} + 1/\kappa_{B}$.

In order to compute the magnetic field in our calculations, we choose the profile described by equation (\ref{Bprof}). It should be noted that the default field profile used here can be appropriately modified to achieve any desired parameterization within the \textsc{stars} code. The numerical routine \textsc{statef.f} receives as input the density at a given mesh point in the model and computes the magnetic field. 
The magnetic contribution to the density is then computed as $\rho_{B} = B^2\,/\,8\pi c^2$ and that to the pressure as $P_{B} = B^2\,/\,8\pi$. The magnetic pressure and density contributions are added to the current model mesh point pressure and density, before the subroutine continues to compute the remaining thermodynamic quantities. The $\kappa_{\rm tab}$ opacity computations are then completed before $\kappa_B$ is computed and added in inverse to the tabulated opacity. This allows for the opacity to be computed in a self-consistent manner rather than switching from Kramers' opacity, or \cite{Itoh1983} style electron conduction opacity, to $\kappa_B$ at specific $B$ and $T$. 

To model strongly magnetised and super-Chandrasekhar WDs, we use the \textsc{stars} code to generate a zero-age main-sequence (ZAMS) star with $M=3\,M_{\odot}$ and $Z = 0.02$. The star is then evolved up to the asymptotic giant branch (AGB) stage until its carbon-oxygen core has grown to about $0.6\,M_{\odot}$. At this point, the chemical evolution of the star is halted and an artificial mass--loss mechanism is enabled to strip the outer envelope until a CO WD with a thin helium atmosphere is formed. The computation of $B$ is then enabled once the model in question has relaxed and allowed to proceed along its cooling track. Setting a particularly large $B$ (either in terms of a large $B_{\rm s}$, a large $B_{0}$ or both) typically requires the field parameters to be increased in stages to allow the model to relax. The residual mass can then be directly added or removed in order to produce a model WD of any desired mass.

\begin{table}
\begin{center}
\caption{The effect of magnetic field $B$ on the numerical mass--radius relation as computed with the \textsc{stars} code for luminosity $L = 10^{-4}\ L_{\odot}$. Here $T_{\rm c}$ and $\rho_{\rm c}$ represent the temperature and density are computed at the central calculation point at the timestep where the model reaches $L = 10^{-4}\ L_{\odot}$ on its cooling curve. 
} 
\label{Table4}
\bgroup
\def\arraystretch{1.1}
\begin{tabular}{| c | c | c | c | c | c | c |}
\hline
\hline
\centering
$M/M_{\odot}$ & $(B_{\rm s},B_0)\, {\rm G}$ & $T_{\rm c}/10^6 \ {\rm K}$ & $\rho_{\rm c}/10^6 \ {\rm g\ cm^{-3}}$ &  $R/1000\ {\rm km}$\\ \hline \hline
0.08 & $(0,0)$ & 5.22 & 0.0242 & 23.307   \\ 
    & $(10^7,10^{12})$ & 5.23 & 0.0242 & 23.342   \\
    & $(10^7,10^{13})$ & 5.28 & 0.0241 & 23.950   \\
    & $(10^6,10^{14})$ & 5.42 & 0.0224 & 24.498   \\\hline
0.15 & $(0,0)$ & 4.17 & 0.1036 & 16.810  \\ 
    & $(10^7,10^{12})$ & 4.17 & 0.1036 & 16.818  \\
    & $(10^7,10^{13})$ & 4.35 & 0.1031 & 16.982   \\
    & $(10^6,10^{14})$ & 4.40 & 0.0957 & 17.395   \\\hline
0.25 & $(0,0)$ & 3.37 & 0.3293 & 13.425 \\ 
    & $(10^7,10^{12})$ & 3.38 & 0.3293 & 13.426   \\
    & $(10^7,10^{13})$ & 3.51 & 0.3285 & 13.475   \\
    & $(10^6,10^{14})$ & 3.55 & 0.2991 & 13.850   \\\hline
0.45 & $(0,0)$ & 2.76 & 1.4330 & 10.240 \\ 
    & $(10^7,10^{12})$ & 2.74 & 1.4330 & 10.239   \\
    & $(10^7,10^{13})$ & 2.84 & 1.4300 & 10.255   \\
    & $(10^6,10^{14})$ & 2.85 & 1.2380 & 10.658   \\\hline
0.62 & $(0,0)$ & 2.46 & 3.7650 & 8.533 \\ 
    & $(10^7,10^{12})$ & 2.47 & 3.7670 & 8.530   \\
    & $(10^7,10^{13})$ & 2.47 & 3.7580 & 8.538   \\
    & $(10^6,10^{14})$ & 2.53 & 3.0420 & 9.012   \\\hline \hline
\end{tabular}
\egroup
\end{center}
\end{table}

\subsection{Model results}
We use the \textsc{stars} evolution code with the modifications described to create a grid of B-WD models with a range of masses and field parameters. 
We use our grid of models to investigate qualitatively the B-WD mass--radius relationship at different fields, with the objective of numerically validating our analytical models. In Table~4, we list the central temperature, central density and radius for each of our fixed mass numerical models and for each set of field configurations. In all cases, the models are allowed to cool until the luminosity has reached $L = 10^{-4}\, L_{\odot}$. Furthermore, the stellar composition is held fixed and equivalent to the description in Section 2. 
Unlike in our analytical models, we do not consider separate core and envelope regions but instead allow for the cooling to occur naturally in the numerical models with no explicit prescription. It is not trivial to produce numerically stable B-WD models with arbitrary field configurations with \textsc{stars}, so we limit the range of field configurations as opposed to our analytical treatment.

In Table 4, we present a number of trends in $T_{\rm c}$, $\rho_{\rm c}$ and $R$ for a range of mass and magnetic field. As expected, the B-WD radius decreases with $M$, as shown in Figure 3 and Table 2. For models of the same mass, $R$ increases very slightly as a function of the magnetic field, until field parameter $B_{0} = 10^{14}\, {\rm G}$ is reached, at which point $R$ increases significantly, consistent with our analytical expectation from Section 3. Here the central density $\rho_{\rm c}$ is not the density in the core but rather the density attained at the central calculation point in our model. This is equivalent to our core and envelope analytical approach described in Tables 2 and 3. The central temperature $T_{\rm c}$ is also computed in a similar manner as in Section 3. As expected, $\rho_{\rm c}$ increases rapidly with an increase in $M$. However, $\rho_c$ drops marginally once the threshold field $B_0 \approx 10^{14}\, {\rm G}$ is reached. This results from a corresponding reduction in $R$. The comparison of $\rho_{\rm c}$ for a given $R$ with the analytical models presented in Table 2 clearly indicates that the central density does in fact increase as the field increases, with a sharp rise in the density as the critical $B_0$ is reached. This is in line with our earlier expectation that $\rho_{\rm c}$ must increase to compensate for the increased magnetic pressure as the field increases.

\begin{figure}
\includegraphics[width=1.12\columnwidth]{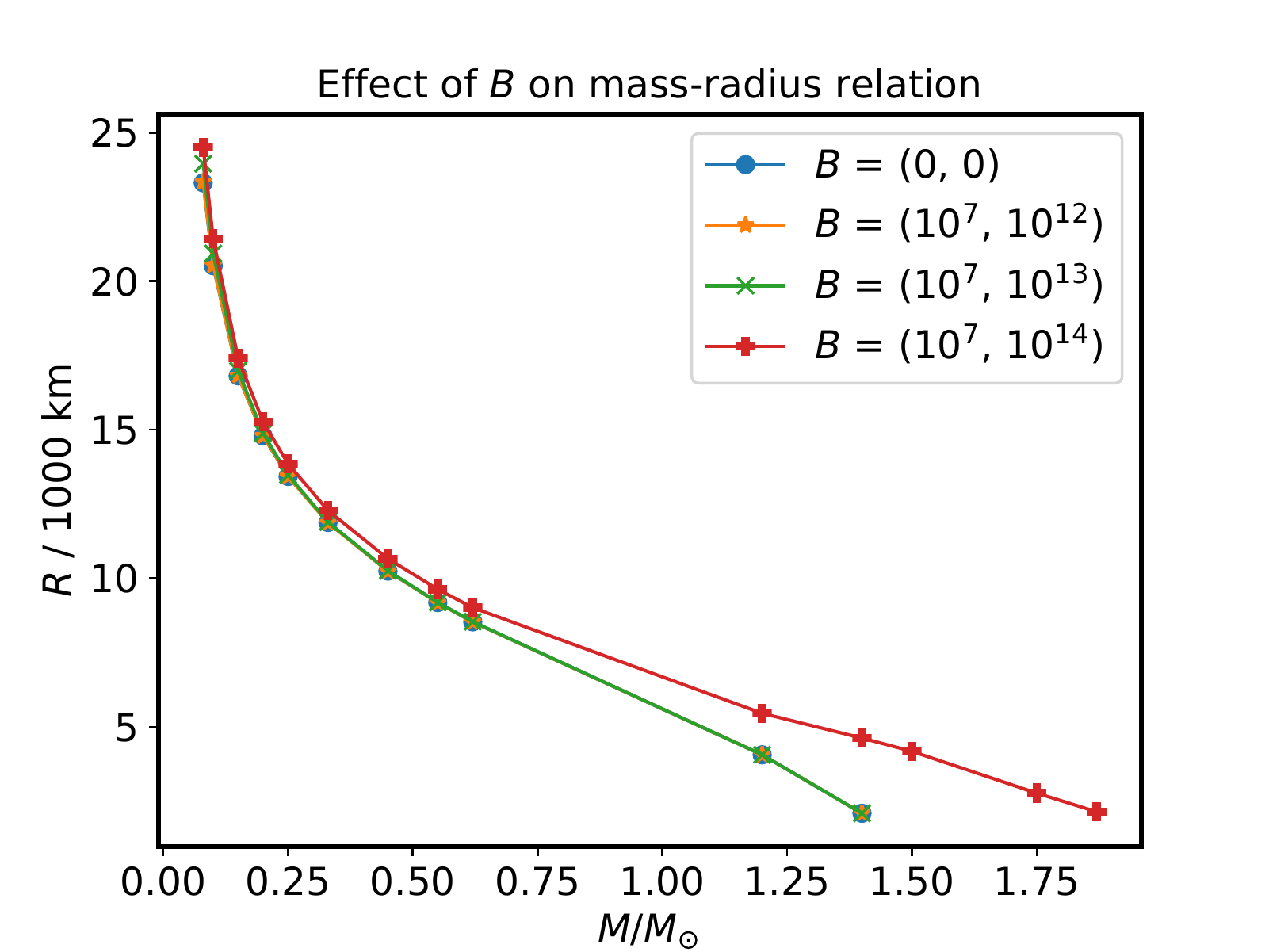}
\caption{The effect of magnetic field on the mass--radius relation of highly magnetized WDs for $B=(0, 0)$ (blue circles), $B=(10^7, 10^{12})\, {\rm G}$ (orange stars), $B=(10^7, 10^{13})\, {\rm G}$ (green crosses) and $B=(10^7, 10^{14})\, {\rm G}$ (red pluses). Unlike in Table \ref{Table4}, we do not compute these numerical models at a fixed luminosity, but rather at the stage where they have already cooled down to the point such that the code fails to converge further.  
}
\label{fig7}
\end{figure}

In Figure 7, we validate the mass--radius relations described by our analytical model in Figure 3. As before, the effect of increasing $B$ on the mass--radius relation is analogous to increasing $L$. This corresponds to an equivalent star earlier on its WD cooling curve. Rather than specifying a WD radius and then inferring its mass, it is more reasonable to base our numerical grid on fixed masses and then compute $R$. 
It should be noted that the numerical models here are not computed for a fixed luminosity but at the time when the models have cooled to the point where the code no longer converges. This is owing to the limitations of the EoS utilised in the \textsc{stars} code at low temperatures and high densities. The radii of these models at low temperatures are almost entirely independent of luminosity. This is expected as the thermal pressure support in all of these models is negligible compared to degeneracy pressure support as well as the magnetic pressure support. The three curves representing $B_{0}=0$, $B_{0}=10^{12}\, {\rm G}$ and $B_{0}=10^{13}\, {\rm G}$ almost completely overlap throughout. This is due to the fact that the magnetic pressure contribution only becomes comparable to the degeneracy pressure contribution for the highest field strength. This reflects the same trend observed at fixed luminosity in Table \ref{Table4}. 

We obtain results that are in good agreement with our analytical formalism and the magnitude of $B_{0}$ dictates the shape of the mass--radius curve. 
For stronger fields, specifically with a larger $B_0$, the mass--radius relation deviates from the zero/low-field relation, with the deviation increasing at larger masses. In particular, for $B=(10^7,10^{14})\, {\rm G}$, we obtain super-Chandrasekhar WDs with limiting mass $\sim 1.9\, M_{\odot}$.
As anticipated, the radii inferred from our numerical models are not exactly equal to those computed analytically. The EoS for our numerical models is computed with the standard solver in \textsc{stars} and this essentially differs from the purely analytical estimates. We also include the effects of neutrino losses in our numerical models. The neutrino cooling effect may cause non-negligible energy losses from the cores of very hot and/or dense WD stars. 
Based on the prescription given by \cite{Itoh1983}, we model the neutrino losses that become significant once $T \geq 10^{7}\, {\rm K}$ and $\rho \geq 10^{10}\, {\rm g\,cm^{-3}}$ in the stellar matter. 
While no $L=10^{-4}\, L_{\odot}$ model is hot enough for these losses to occur, as listed in Table 4, many of these models would have had sufficiently hot cores at other point on their cooling curves for neutrino losses to occur.

\begin{table}
\begin{center}
\caption{The maximum attainable B-WD mass is computed for the \textsc{stars} models as a function of magnetic field parameters ($B_{\rm s}$,~$B_{0}$). The central density and B-WD radius are listed for the corresponding cases.
}
\def\arraystretch{1.1}
\begin{tabular}{| c | c | c | c |}
\hline \hline
($B_{s}$, $B_{0}$)/G & $\rho_{c}/10^{6}\, {\rm g\,cm^{-3}}$ & $R/1000\, {\rm km}$ & Max Mass/$M_{\odot}$ 
\\ \hline \hline
(0, 0)               &2210 &2.1177 & 1.4397                                                     \\
($10^7$, $10^{11}$)  &2257 &2.1196 & 1.4358                                                     \\
($10^7$, $10^{12}$)  &2280 &2.1227 & 1.4358                                                     \\
($10^7$, $10^{13}$)  &2295 &2.1240 & 1.4373                                                     \\
($10^7$, $10^{14}$)  &2260 &2.1412 & 1.8703                                                     \\\hline \hline
\end{tabular}
\end{center}
\label{tab5}
\end{table}

\begin{figure*}
	\includegraphics[width=0.49\textwidth]{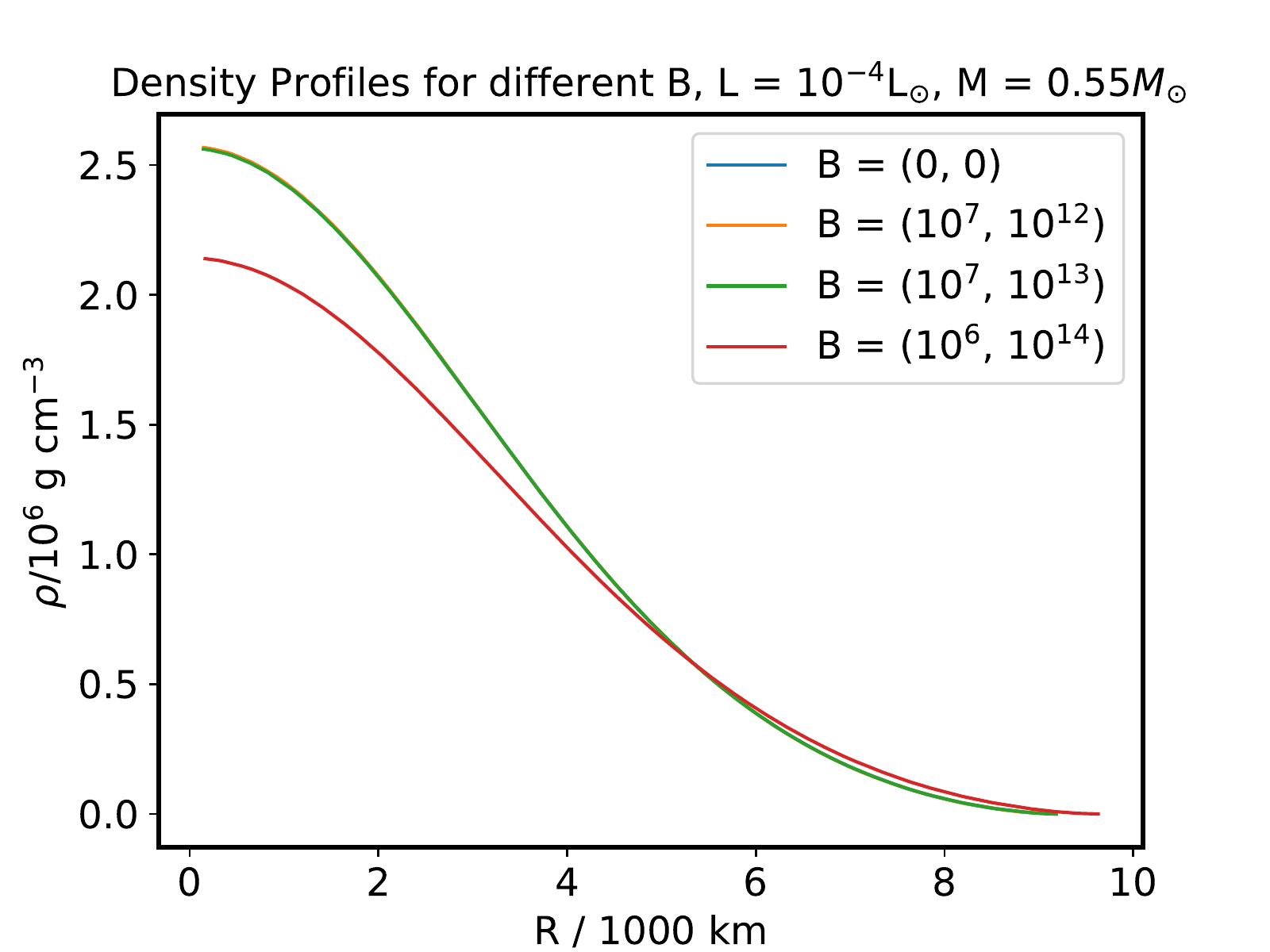}
	\includegraphics[width=0.49\textwidth]{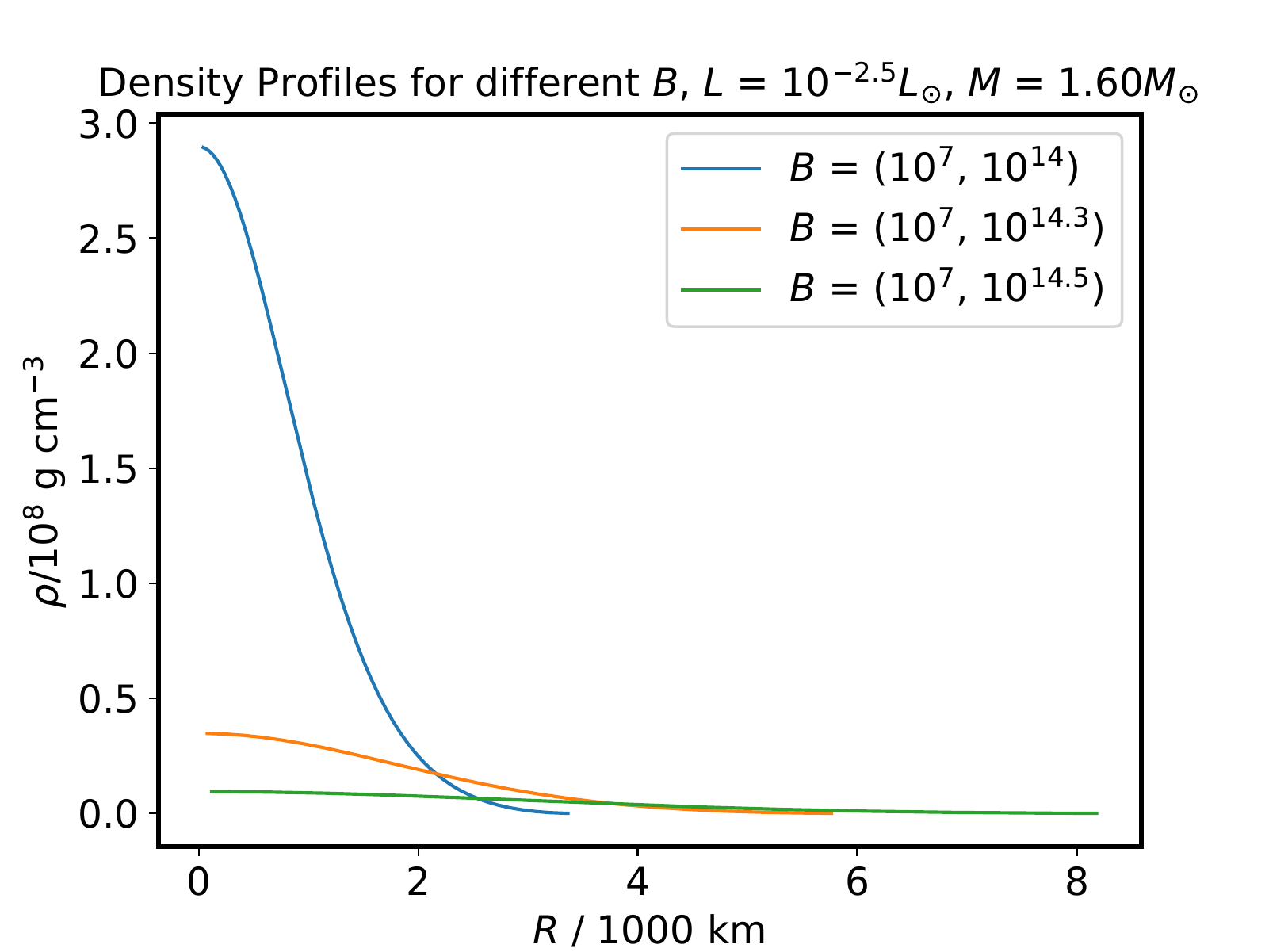}
    \caption{
    {\bf Left panel:} The variation of density as a function of radius is shown for B-WDs with varying magnetic fields $B=(0,0)\ {\rm G}$ (blue), $B=(10^7,10^{12})\ {\rm G}$ (orange), $B=(10^7,10^{13})\ {\rm G}$ (green) and $B=(10^6,10^{14})\ {\rm G}$ (red). Each model has a mass $M=0.55\, M_{\odot}$ and has been allowed to cool to a luminosity of $L=10^{-4}\, L_{\odot}$. This model is essentially equivalent to those computed analytically in Figure 4. 
    {\bf Right panel:} The variation of density as a function of radius is shown as in the left panel, but for a super-Chandrasekhar model mass of $M=1.6\, M_{\odot}$ with varying magnetic fields $B=(10^7,10^{14})\ {\rm G}$ (blue), $B=(10^7,10^{14.3})\ {\rm G}$ (orange) and $B=(10^7,10^{14.5})\ {\rm G}$ (green). Each model has been allowed to cool to a luminosity of $L=10^{-2.5}\, L_{\odot}$. This larger luminosity was required as a result of the simulation's EoS encountering difficulties at lower temperatures. Naturally, a certain minimum value of $B_{0}$ is required to produce a super-Chandrasekhar model, hence we have no models corresponding to the lower field models in the $0.55\,M_{\odot}$ model.
    }
\label{fig8}
\end{figure*}

\begin{figure*}
    \includegraphics[width=0.49\textwidth]{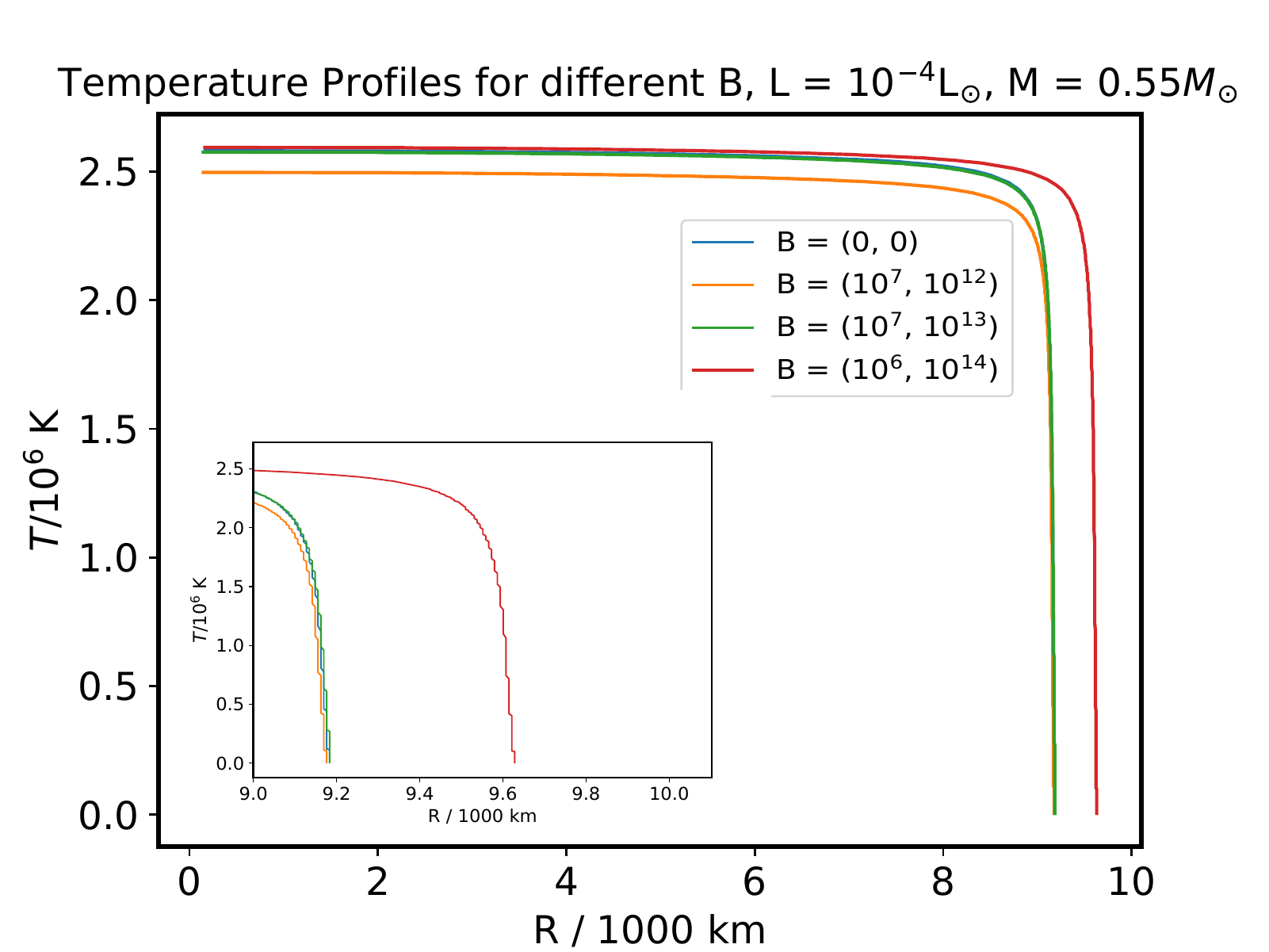}
    \includegraphics[width=0.49\textwidth]{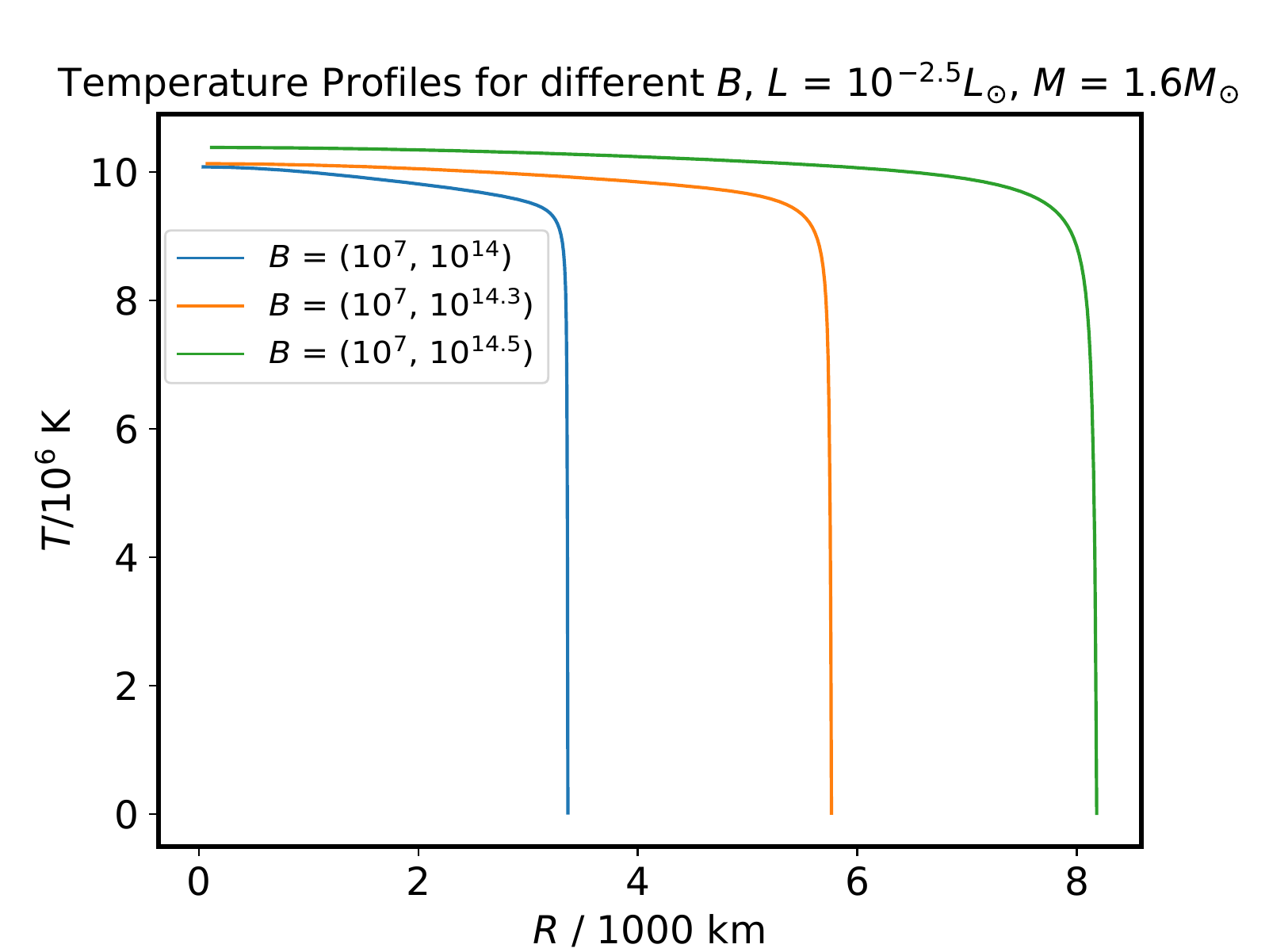}
    \caption{
    {\bf Left panel:} The variation of temperature as a function of radius is shown for B-WDs with varying magnetic fields $B=(0,0)\ {\rm G}$ (blue), $B=(10^7,10^{12})\ {\rm G}$ (orange), $B=(10^7,10^{13})\ {\rm G}$ (green) and $B=(10^6,10^{14})\ {\rm G}$ (red). These models are essentially equivalent to the analytical models presented in Figure 4.
    The mass is fixed at $M=0.55\, M_{\odot}$ for each model, so the radius varies as a function of the magnetic field, as in Figure 7. For each field configuration, the model has been allowed to cool until it reaches $L=10^{-4}\, L_{\odot}$. 
    {\bf Right panel:} The variation of temperature as a function of radius is shown as in the left panel, but for a super-Chandrasekhar model mass of $M=1.6\, M_{\odot}$ with varying magnetic fields $B=(10^7,10^{14})\ {\rm G}$ (blue), $B=(10^7,10^{14.3})\ {\rm G}$ (orange) and $B=(10^7,10^{14.5})\ {\rm G}$ (green). As in Figure \ref{fig8}, each model has been allowed to cool to a luminosity of $L=10^{-2.5}\, L_{\odot}$.
    }
    \label{fig9}
\end{figure*}

Next, we investigate the results presented in Section 3.3, which suggest the possibility of obtaining super-Chandrasekhar WDs provided that the central magnetic field and $B_{0}$ are sufficiently large. Using our modified \textsc{stars} code, we compute the highest stable mass model for a range of field configurations. Our results are summarised in Table 5 and are consistent with the Chandrasekhar mass limit being retained for $B_0 \lesssim 10^{13}\, {\rm G}$, while allowing for the existence of super-Chandrasekhar B-WD models for larger $B_0$. 
We consider a surface field $B_{\rm s} = 10^{7}\, {\rm G}$ for the WD models with $B_{0} = 10^{14}\, {\rm G}$. Regardless, the limiting mass obtained for $(B_{\rm s}, B_{0})=(10^{7},10^{14})\, {\rm G}$ with \textsc{stars} is in perfect agreement with $M \approx 1.865\, M_{\odot}$ for $(B_{s}, B_{0}) \approx (10^{7-9},10^{14})\, {\rm G}$ from the analytical results presented in Table 2. This supports our earlier finding that $B_{\rm s}$ has no appreciable effect on the mass--radius relation. 

Table 2 and the left panel of Figure 4 have shown the results for analytical computation of the density variation as a function of B-WD radius. As stated, we expect the central density for WDs with stronger fields to be larger in order to compensate for the additional magnetic pressure. Figure 8 shows the numerical validation of the same trend for two mass models, i.e., for $M = 0.55\, M_{\odot}$ and $M = 1.6\, M_{\odot}$, with varying field configurations.  
At first sight, the numerical results appear to be inconsistent with our earlier analytical prediction that $\rho_{\rm c}$ should increase to compensate for increased $P_B$ as field increases. However, it should be noted that the analytical computations are performed with a fixed radius, whereas our equivalent numerical models are generated assuming a fixed mass. For the super-Chandrasekhar white dwarf models in the right panel of Figure \ref{fig8} with $M = 1.6\, M_{\odot}$, we cannot investigate the lower field cases as in the left panel of that figure as there are no solutions for lower values of $B_{0}$, hence we elect to investigate field configurations with higher values of $B_{0}$ instead, in an attempt to elucidate a trend in density profiles for these models. 
We find that for $M = 1.6\, M_{\odot}$ and $L = 10^{-2.5} \, L_{\odot}$ the radius and density profiles of the model are very strongly dependant on the value of $B_{0}$. As $B_0$ increases from $10^{14} \, {\rm G}$ to $10^{14.3} \, {\rm G}$ to $10^{14.5} \, {\rm G}$, the radius of the model expands from $\approx$ $3500 \, \rm km$ to $\approx$ $6000 \, \rm km$ to $\approx$ $8000 \, \rm km$, while the central density falls considerably from $\approx$ $3\, \times \, 10^{8}\, \rm g\, cm^{-3}$ to $\approx$ $2 \, \times \, 10^{7} \, \rm g\, cm^{-3}$ over the range of values considered for $B_0$. This is good confirmation that for these models, the total pressure is dominated by the degeneracy pressure and magnetic pressure, with thermal support being negligible, as expected, and that the density structure and radii of the models are largely functions of the value of $B_0$ alone. 
Hence, a model's radius is a function of its magnetic field parameters as well as its mass at a fixed luminosity. Therefore, the $M=0.55\, M_{\odot}$ model with a larger central field has a larger radius and hence a lower mean density. If we compare the central density of the model, rather than just the relative order, our numerical results are indeed consistent with the analytical results for $R=10000\, {\rm km}$ WDs, with central density $\rho_{\rm c} \approx 2.2\times 10^{6}\, {\rm g\, cm^{-3}}$ obtained for $B=(10^{7}, 10^{14})\, {\rm G}$. 

In relation to the results presented in the right panel of Figure 4, we show the variation of the temperature as a function of the radius in Figure 9. In particular, we choose the same mass models and magnetic field configurations as in Figure 8. The analytical model have suggested that the core temperature is primarily determined by the luminosity and is largely unchanged with variation in magnetic field. In good agreement with this prediction, we find here that the core temperatures of models with masses $M = 0.55\, M_{\odot}$ and $M = 1.6\, M_{\odot}$ are in fact largely unchanged with varying magnetic field. The small difference in central temperatures with magnetic field between these numerical models is a result of the difference in radius and hence in the mean density across the models. In the super-Chandrasekhar case however, the radii, central temperatures and central densities of the models are very highly dependant on the field configuration. This is, in part, a consequence of the pressure support no longer being dominated by degeneracy pressure, but rather both magnetic and degeneracy pressures. Hence, we observe that in the super-Chandrasekhar case the density and temperature profiles, as well as the radii of the models are highly dependant on the field structure. Our 
numerical models further demonstrate that the radial temperature gradient $dT/dr$ within the surface layers of each model falls as the magnetic field increases.

\subsection{Summary of simulation results}
We have produced a novel set of modifications to the \textsc{stars} code based on our magnetic field prescription (see equation~\ref{Bprof}) in order to compute a grid of numerical models of highly magnetised WDs. This methodology has allowed us to validate qualitatively the analytical results, from Section 3 in particular. A qualitative approach is necessary because we cannot trivially fix the B-WD radius within the \textsc{stars} framework, and are rather restricted to fixing the mass to generate models that are essentially analogous to our analytical models. 
We have determined, as we had analytically, that the effect of the surface field $B_{\rm s}$ is not significant, while the central magnetic field and hence $B_{0}$ can significantly affect the B-WD mass--radius relation. 
This is confirmed because the mass $M \approx 1.87\, M_{\odot}$ computed analytically for $(B_{\rm s}, B_{0}) = (10^{9}, 10^{14})\, {\rm G}$, is in very close agreement with the mass $M \approx 1.89\, M_{\odot}$ inferred from our numerical models for $(B_{\rm s}, B_{0}) = (10^{7}, 10^{14})\, {\rm G}$, i.e. despite the two orders of magnitude difference in the surface magnetic fields between the two cases. In Table 5, we have demonstrated that stable numerical models of highly magnetised super-Chandrasekhar WDs can be created provided that the central magnetic field is sufficiently large.

\section{Summary \& Conclusions}
Sufficiently strong magnetic fields can alter the EoS of electron degenerate matter to yield super-Chandrasekhar B-WDs with masses that can be as high as $M \approx 2.6\, M_{\odot}$, even in the absence of rapid rotation (see \citealt{DM2012,DM2013,SM2015}). Besides elevating the limiting mass of WDs, strong fields can also impact the thermal characteristics of the underlying compact star and thereby its observed properties. \citet{MBh2018} studied the luminosity suppression in B-WDs with $B \gtrsim 10^{12}\, {\rm G}$, assuming that the interface properties are essentially similar to their non-magnetised counterparts, to demonstrate that their cooling rates are significantly attenuated for such strong magnetic fields. Subsequently, \citet{Gupta2020} improved on this preliminary analytical model by removing the assumption of fixed interface parameters that was initially considered for WDs with a preassigned mass. They further derived the mass--radius relation as well as the stellar structure properties within both non-degenerate and degenerate regions of the B-WD.  

In this paper, we have revisited the physics of luminosity suppression and its effect on the mass--radius relation for highly magnetised WDs. We have included the contributions from the electron-degenerate isothermal core, ideal gas surface layer and magnetic field to model the B-WD structure properties by solving the magnetostatic equilibrium, photon diffusion and mass conservation equations. In order to distinguish the strongly magnetised cases from the weakly/non-magnetised cases, we have appropriately amended our treatment of the radiative opacity, magnetostatic pressure balance and EoS for the degenerate core. Although an increase in surface luminosity results in higher total mass, especially for larger WDs, we have shown that the Chandrasekhar mass limit is retained for $10^{-4} \leq L/L_{\odot} \leq 10^{-2}$. The increase in luminosity for a given stellar radius leads to a larger $\rho_{\rm c}$ and more compact degenerate interior and so an increased capacity to hold more mass. Although $B_{\rm s}$ has negligible effect on the B-WD mass, $B_0$ affects the shape of the mass--radius relation by shifting the curve towards higher masses for stronger fields for a given radius. In particular, strong fields with $B_0 \approx 10^{14}\, {\rm G}$ can raise $\rho_{\rm c}$ significantly and yield super-Chandrasekhar WDs with masses as high as $\sim 1.87\, M_{\odot}$.

We have computed the B-WD luminosity necessary in order to obtain mass--radius relations similar to those for non-magnetised WDs. Provided that the gravitational energy does not vary significantly, an increase in magnetic energy density needs to be compensated by a corresponding reduction in thermal energy, in order to maintain the structural stability of the B-WD. 
Nevertheless we have shown that, even with a significant reduction in their luminosities, the masses of smaller radii B-WDs are not similar to their non-magnetised counterparts. 
In particular, for stars with radii $2000\, \leq R/{\rm km} \leq 6000$, the inferred B-WD mass remains considerably larger leading to an extended branch in the mass--radius relation, even for highly suppressed luminosities such that $10^{-16} \leq L/L_{\odot} \leq 10^{-12}$. To model the time variation of the B-WD structure properties, we have also considered the cooling evolution of these stars in the presence of strong field dissipation processes, particularly Ohmic dissipation and Hall drift. As the star gradually evolves over time, its thermal energy is radiated away in the observed luminosity from the surface layers. Owing to primarily field decay and also simultaneous cooling over the typical WD age $\tau \approx 10\, {\rm Gyr}$, the luminosity adjusted masses turned out to be significantly lower than their initial estimates. Even though the limiting B-WD mass is lowered significantly to $1.5\, M_{\odot}$ compared to $1.9\, M_{\odot}$ at initial time, and corresponding initial field $B=(10^9,10^{14})\, {\rm G}$, the majority of these systems still remain practically hidden throughout their cooling evolution due to their strong fields and consequently low luminosities. 

We have also explored a set of stellar evolution models for B-WDs using a modified version of the \textsc{stars} code with the objective of numerically validating our analytical formalism. For this, we have appropriately modelled the chemical evolution of the star along its cooling track and included the effects of energy losses due to neutrino cooling from the cores of very hot and/or dense WDs. In validation of our analytical approach, we have found that the limiting mass $\sim 1.8703\, M_{\odot}$ obtained with the \textsc{stars} numerical models is in very good agreement with $M \approx 1.87\, M_{\odot}$ inferred from the analytical calculations for WDs with strong fields $B=(10^{6-9},10^{14})\, {\rm G}$. However, the results presented in this work argue that the young super-Chandrasekhar B-WDs may not sustain very large masses over the course of their entire cooling evolution, and this essentially explains their apparent scarcity even without the difficulty of detection owing to their suppressed luminosities. 

We have adopted a considerably more generalised framework to improve upon the analytical approach presented in our previous studies \citep{MBh2018,Gupta2020}. In addition to radiative cooling, we have now incorporated the effect of convective energy transport through the photon diffusion equation. To account for the effects of strong magnetic fields on the magnetostatic pressure balance and Landau quantised electron energy states, we have considered the general relativistic TOV equation and appropriately modified the interface condition.  
Furthermore, we have included typical neutrino cooling energy losses with \textsc{stars} as well as the dissipation of strong magnetic fields over time. 

We have nevertheless made some simplifying physical assumptions in order to improve our modelling of the B-WD structure properties. Firstly, we have assumed that the B-WDs are approximately spherical for $B_0 \lesssim 10^{14}\, {\rm G}$, as demonstrated by \citet{SM2015} for toroidally dominated fields. Secondly, we have adopted a constant radial luminosity profile $L_r=L$ for our models as no H burning or other nuclear fusion reactions occur within the degenerate WD core. Thirdly, we have used a temperature independent EoS for the electron degenerate pressure (see \citealt{Gupta2020} for physical justification) and uniform B-WD core temperature (only for the analytical model) due to the high thermal conductivity of the degenerate electrons. Lastly, we have assumed the self-similarity of the cooling process for evolution over the typical WD age of $\tau \sim 10\, {\rm Gyr}$. Future work can address the effect of more complicated magnetic field geometry on the inferred mass--radius relation as well as the effect of varying stellar compositions on the B-WD cooling evolution.

\section*{Acknowledgments}
M.B. acknowledges support from Eberly Research Fellowship at the Pennsylvania State University. A.J.H. thanks the Science and Technology Facilities Council (STFC) and the Cambridge Commonwealth, European \& International Trust for his doctoral funding.
C.A.T. thanks Churchill College for his fellowship. B.M. acknowledges a partial support by a project of Department of Science and Technology (DST-SERB) with research Grant
No. DSTO/PPH/BMP/1946 (EMR/2017/001226).

\section*{Data availability}
The data produced in this study will be shared on reasonable request to the authors.

\label{lastpage}

\end{document}